\documentclass[11pt]{article}
\usepackage[preprint]{acl}
\usepackage{times}
\usepackage{latexsym}
\usepackage[T1]{fontenc}
\usepackage[utf8]{inputenc}
\usepackage{microtype}
\usepackage{inconsolata}
\usepackage{graphicx}
\usepackage{amsmath}
\usepackage{hyperref}
\usepackage[nameinlink]{cleveref} 
\usepackage{algorithm}
\usepackage{algorithmic}
\usepackage{url}
\usepackage{xurl}
\usepackage{colortbl}
\usepackage{booktabs}
\usepackage{multirow}
\usepackage{subcaption}
\usepackage{adjustbox}
\usepackage{tikz}
\usepackage[most]{tcolorbox}
\usepackage{xcolor}
\usepackage{caption}
\usepackage{float}
\usepackage{amssymb}

\usepackage{seqsplit}
\newcommand{\HeaderCell}[1]{\textbf{#1}}
\newcommand{\StructCell}[1]{#1}
\newcommand{\EBNFRule}[2]{%
  \parbox[t]{0.72\columnwidth}{\texttt{#1}\\#2}%
}

\definecolor{codebg}{HTML}{F8FAFC}        
\definecolor{framegray}{HTML}{CBD5E1}     
\definecolor{titlegray}{HTML}{334155}     
\definecolor{highlightblue}{HTML}{DBEAFE} 
\definecolor{commentgreen}{HTML}{15803D}
\definecolor{keywordblue}{HTML}{1D4ED8}
\definecolor{stringred}{HTML}{B91C1C}
\definecolor{numbergray}{HTML}{64748B}

\tcbset{
    codebox/.style={
        enhanced,
        colback=codebg,
        colframe=framegray,
        boxrule=0.5pt,
        arc=2pt,
        left=2pt,
        right=2pt,
        top=2pt,
        bottom=2pt,
        boxsep=2pt,
        width=\columnwidth,
        before skip=4pt,
        after skip=4pt,
    }
}
\title{Preserving Structure, Strengthening Signals: A Grammar-Driven Approach to Optimize the Trade-off Between Quality and Detectability in Code Watermarking}
\author{
  \textbf{Licheng Yu}\textsuperscript{1},
  \textbf{Aiwei Liu}\textsuperscript{2}, 
  \textbf{Songze Li}\textsuperscript{1}\thanks{Corresponding author.}
\\
  \textsuperscript{1}Southeast University  \quad
  \textsuperscript{2}Tsinghua University
\\
  \small{\texttt{\{lichengyu, songzeli\}@seu.edu.cn, liuaiwei20@gmail.com}}
}

\begin{document}
\maketitle
\begin{abstract}
Text watermarking has emerged as a crucial technique for identifying AI-generated content. However, current watermarking schemes entail a trade-off between text quality and watermark detectability, which is further intensified by the inherent low-entropy nature of code. To mitigate this dilemma, we propose \textbf{G}rammar-\textbf{D}riven \textbf{W}atermark (\textbf{GDW}), a training-free watermarking approach tailored to LLM-based code generation. GDW employs a three-level masking mechanism consisting of a green-list mask, a syntactic validity mask, and a structural role mask to regulate watermark injection. These masks are integrated through hierarchical modulation, which limits the scope of watermark injection to syntactically admissible green-list tokens and adjusts watermark strength based on structural roles through a role function. Accordingly, we design a role-aware weighted detection statistic. For comprehensive evaluation, we introduce the \textbf{A}rea \textbf{U}nder the \textbf{T}rade-off \textbf{C}urve (\textbf{AUTC}), a metric that quantifies the quality-detectability trade-off. Extensive experiments demonstrate that GDW outperforms existing baselines and achieves state-of-the-art (SOTA) trade-off performance, while maintaining superior performance under seven semantics-preserving attacks.
\end{abstract}

\section{Introduction}

In recent years, large language models (LLMs) have become indispensable for programming assistance, powering widely used tools such as Cursor \citep{Cursor} and Claude Code \citep{ClaudeCode}. However, as LLM-generated code is increasingly integrated into commercial software, concerns over code provenance and potential copyright infringement have increased substantially \citep{dey2019software}. Therefore, algorithms that can effectively identify machine-generated code have become crucial.

\begin{figure}[ht]
\centering
  \includegraphics[width=1.0\linewidth]{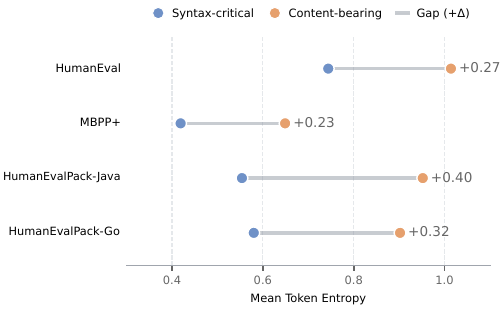}
  \caption{Mean token entropy across benchmarks, measured using Qwen2.5-Coder-3B. We categorize tokens into syntax-critical and content-bearing categories according to their structural roles. Content-bearing tokens consistently exhibit higher entropy, motivating role-aware hierarchical watermark modulation.}
  \label{fig:Mean_Token_Entropy_Comp}
  \vspace{-6mm}
\end{figure}

Watermarking has emerged as a promising solution by embedding imperceptible but statistically detectable signals into model outputs. One dominant paradigm is logits-based watermarking, which pseudorandomly partitions the vocabulary into a ‘green list’ and a ‘red list’, and then increases the logits of green-list tokens \citep{kirchenbauer2023watermark}. However, while logits-based watermarking methods have proven effective for natural language generation, directly applying this paradigm to code generation remains challenging. Programming languages are inherently low-entropy, as rigid syntactic rules heavily concentrate output probabilities onto a narrow candidate set. In such low-entropy settings, naively injecting watermark bias forces a stringent trade-off between code quality and detectability: a strong bias risks distorting structurally critical tokens and degrading functional correctness, whereas a weak bias fails to accumulate sufficient statistical evidence for reliable detection.

Existing methods mainly alleviate this tension from two perspectives. Entropy-based approaches \citep{lee2024wrote, lu2024entropy} exploit token entropy either to avoid perturbations at low-entropy positions during injection or to enhance detection through entropy weighting. However, token entropy only measures the uncertainty of a model’s prediction at a specific token position, leaving the roles of tokens in program structure underexplored. Recent structure-aware methods either rely on an additional trained predictor \citep{guan2024codeip} or statically exclude syntax tokens from watermark injection \citep{kim2026marking}. While these methods highlight the importance of code structure, their coarse-grained use of structural information limits their ability to balance code quality and watermark detectability. These limitations suggest that effective code watermarking requires a more fine-grained understanding of code structure, so that watermark perturbations do not favor syntactically inadmissible continuations while accumulating sufficient statistical evidence for reliable detection.

Motivated by the need for finer-grained structural modeling, we decompose code watermark injection into two complementary decisions: \emph{where} watermark bias should be applied and \emph{how strongly} different tokens should be biased. For the former, grammatical constraints identify syntactically admissible continuations and prevent watermark perturbations from favoring invalid generation paths. For the latter, structural roles provide a principled basis for assigning different watermark strengths. Figure~\ref{fig:Mean_Token_Entropy_Comp} provides empirical support for this design: syntax-critical tokens warrant conservative modulation, whereas content-bearing tokens offer greater flexibility for strengthening watermark signals.

In this paper, we propose \textbf{G}rammar-\textbf{D}riven \textbf{W}atermark (\textbf{GDW}), a training-free watermarking approach for code generation. GDW incorporates Context-Free Grammar (CFG) constraints into decoding through incremental parsing, restricting watermark bias to syntactically admissible continuations. At each decoding step, GDW constructs a three-level masking mechanism consisting of a green-list mask, a syntactic validity mask and a structural role mask. The first two masks determine where to inject the watermark bias, while the structural role mask controls how strongly different tokens are biased. GDW integrates these masks through unified hierarchical modulation, assigning conservative bias to syntax-critical tokens and stronger bias to content-bearing tokens. Detection is performed with a role-aware weighted statistic aligned with the injection strategy.

For comprehensive evaluation, we introduce the Area Under the Trade-off Curve (AUTC), which quantifies the quality-detectability trade-off over a common quality interval. Extensive experiments show that GDW achieves the highest AUTC among all compared baselines. Furthermore, GDW maintains superior trade-off performance under seven semantics-preserving attacks.

Our contributions can be summarized as follows:

\begin{itemize}
    \vspace{-2mm}
    \item We propose \textbf{GDW}, a grammar-driven code watermarking approach that operates entirely at inference time without training or fine-tuning, supporting multiple programming languages, including Python, Java, and Go.

    \vspace{-1mm}
    \item We construct a precomputable grammar-constrained masking mechanism that avoids promoting syntactically invalid tokens during watermark injection, while enabling structural role-aware hierarchical modulation, and design an aligned role-weighted detection statistic. Furthermore, we introduce AUTC as a unified metric for evaluating the quality-detectability trade-off.

    \vspace{-1mm}
    \item Our extensive experiments demonstrate that the GDW approach achieves state-of-the-art (SOTA) performance in terms of the quality-detectability trade-off.
\end{itemize}

\begin{figure*}[t!]
\centering
\includegraphics[width=1.0\linewidth]{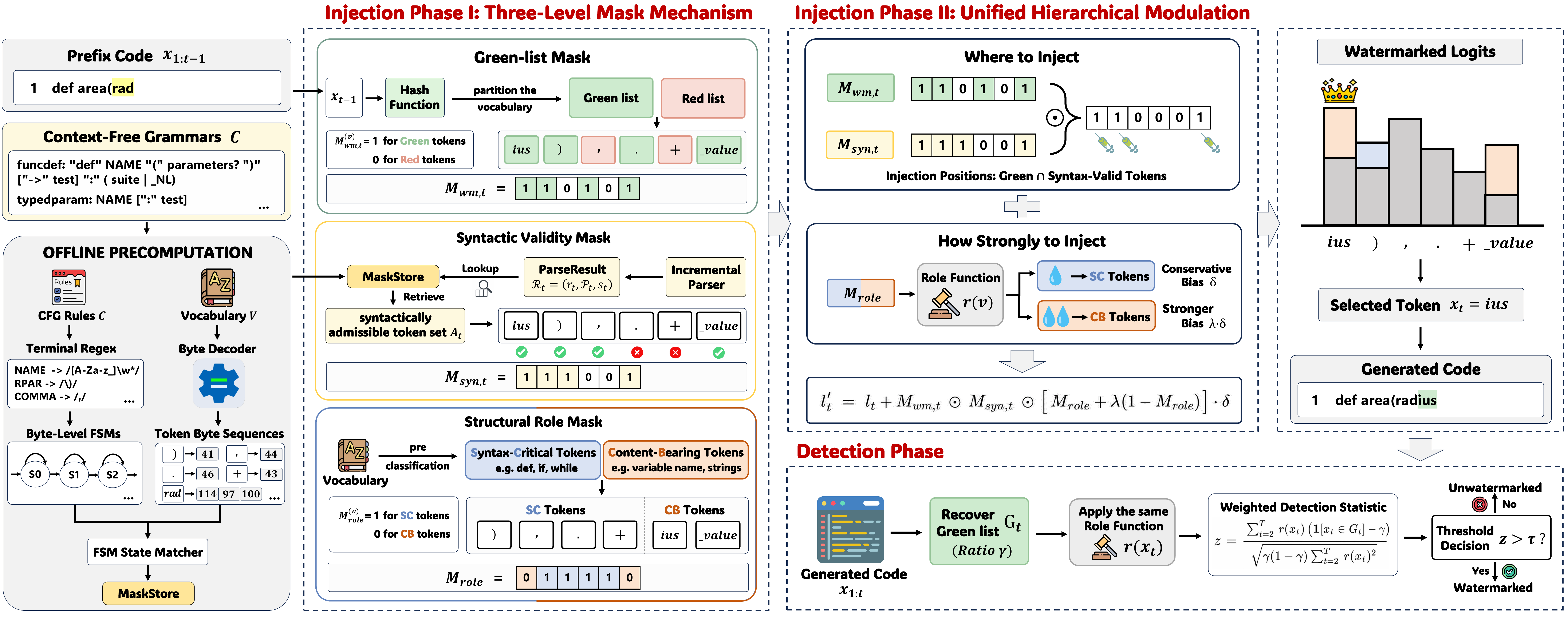}
\caption{Overview of the proposed Grammar-Driven Watermark (GDW) framework. Without syntactic filtering, direct watermark injection may favor a syntactically invalid green-list token, such as the period token (\texttt{.}) in the illustrated example. To address this issue, GDW first performs offline precomputation, in which CFG terminals are compiled into byte-level FSMs and matched against vocabulary-token byte sequences to construct a reusable \textsc{MaskStore}. The \textsc{MaskStore} enables efficient retrieval of syntactically admissible tokens to construct the syntactic validity mask. During decoding, the green-list mask and syntactic validity mask determine watermarkable tokens, while the structural role mask modulates the watermark strength through the role function.}

\label{fig:pipeline}
\vspace{-3mm}
\end{figure*}

\vspace{-4mm}
\section{Related Work}
\paragraph{Text Watermarking for LLMs.}
Watermarking in LLMs refers to the process of embedding imperceptible signals into generated text to identify its source. Text watermarking methods for LLMs have been widely studied for intellectual property protection \citep{li2023functionmarker, li2023protecting, yao2024promptcare}, content provenance and authentication \citep{wang2023towards, sato2023embarrassingly, takezawa2023necessary, liu2024unforgeable, feng2025bimark}, and improving watermark robustness against text modifications \citep{zhao2023provable, kirchenbauer2023reliability, zhang2024remark, liu2024semantic, ren2024robust, qu2025provably}. Despite substantial progress in natural-language text \citep{huo2024token, mao2024watermarking, wang2025morphmark}, a key challenge remains in optimizing the trade-off between watermark detectability and the quality of generated code.

\paragraph{Code Watermarking.}
Early works on static code watermarking focused on program structures or binary representations \citep{collberg1999software, collberg2005software, balachandran2014function, chen2017software}, but proved brittle for LLM-generated code. Researchers have explored selecting watermark embedding positions through token entropy. SWEET \citep{lee2024wrote} filters tokens by entropy threshold, while EWD \citep{lu2024entropy} employs entropy-weighted detection. Although these methods mitigate quality degradation, their rigid exclusion of low-entropy tokens leaves room for further improvement. Recent studies have improved watermarking methods from the perspective of code structure. \citet{guan2024codeip} introduce a grammar-guided watermarking method that trains a type predictor to identify injection positions. However, this method demands extra computational resources, which reduces its flexibility for large-scale deployment. Meanwhile, \citet{kim2026marking} restrict watermark embedding to non-syntax tokens. In contrast, our Grammar-Driven Watermark moves beyond this static exclusion scheme. 

\section{Preliminaries}
\subsection{LLM Text Generation}
Consider an autoregressive large language model $\mathcal{M}$ with a vocabulary $\mathcal{V}$. Given a prompt or prefix sequence $\mathbf{x}_{1:t-1}=(x_1,\dots,x_{t-1})$, the LLM generates a logit vector $\mathbf{l}_{t} \in \mathbb{R}^{|\mathcal{V}|}$. The next-token probability distribution $P_{\mathrm{LM}}(\cdot\mid\mathbf{x}_{1:t-1})$ over the vocabulary $\mathcal{V}$ is obtained by applying the softmax function to $\mathbf{l}_{t}$. Specifically, the probability of selecting the $k$-th vocabulary token as the next token $x_t$ is given by $p_{k}^{(t)} = {\exp(l_{t,k})}/{\sum_{i=1}^{|\mathcal{V}|} \exp(l_{t,i})}$. During inference, the next token is selected from $P_{\mathrm{LM}}$ using a decoding strategy, such as top-$p$ sampling or beam search. This process is repeated autoregressively by appending each generated token to the prefix until a termination criterion is met.

\subsection{LLM Text Watermarking}
The general watermarking scheme used in this paper is called KGW \cite{kirchenbauer2023watermark}. Given a prefix sequence $\mathbf{x}_{1:t-1}$, the watermark process can be described as follows. At each decoding step, a pseudorandom number generator is seeded using the hash of the preceding token $x_{t-1}$ to generate a random permutation of $\mathcal{V}$. The vocabulary is then split into a green list $G_t$ of size $\gamma |\mathcal{V}|$ and a red list $R_t$ of size $(1-\gamma) |\mathcal{V}|$, where $\gamma \in (0, 1)$ is the green-list ratio. Watermark signals are injected by adding a bias $\delta$ to the logits of tokens in ${G}_t$, making them more likely to be selected during generation. The detection process is formulated as a hypothesis test. Specifically, the detector computes a z-score $z = (|s|_G - \gamma T) / \sqrt{\gamma(1-\gamma)T}$, where $|s|_G$ is the number of green-list tokens in a text of length $T$. The text is identified as watermarked when the z-score exceeds a predefined threshold.

\section{Methodology}
\subsection{Overview}

GDW is motivated by two observations. First, watermark perturbations should not favor continuations that violate the grammatical structure of the target programming language. Second, syntactically admissible tokens differ in their sensitivity to logit perturbations: syntax-critical tokens determine program structure, whereas content-bearing tokens generally allow greater lexical flexibility. Accordingly, GDW decomposes watermark injection into two decisions at each decoding step: \emph{where} and \emph{how strongly} the watermark signal should be injected. During injection, these decisions are implemented through a three-level masking mechanism and a unified hierarchical modulation strategy. For detection, GDW employs a weighted statistic aligned with the injection strategy. Figure~\ref{fig:pipeline} illustrates the overall framework.

\subsection{Grammar-Driven Watermark Injection}
\label{Grammar-Driven Watermark Injection}

In this section, we describe the watermark injection procedure of GDW. At each decoding step, GDW constructs three binary masks over the vocabulary: a green-list mask, a syntactic validity mask, and a structural role mask. The first two determine which tokens can receive watermark bias, while the third distinguishes tokens by structural role to modulate the bias. Together, they enable GDW to control both the scope and strength of watermark injection.

\subsubsection{Green-list Mask.} 
Following the KGW framework \citep{kirchenbauer2023watermark}, GDW pseudorandomly partitions the vocabulary $\mathcal{V}$ into a green list $G_t$ and a red list $R_t$ at each decoding step $t$, based on a hash of the preceding token $x_{t-1}$. We define the green-list mask $M_{\mathrm{wm},t} \in \{0,1\}^{|\mathcal{V}|}$ as:

\vspace{-1mm}
\begin{equation}
M_{\mathrm{wm},t}^{(v)}
=
\begin{cases}
1, & v \in G_t,\\
0, & v \in R_t.
\end{cases}
\label{eq:wm_mask}
\end{equation}

\subsubsection{Syntactic Validity Mask.} 

Syntactically invalid tokens can disrupt program structure and degrade code quality. To prevent watermark bias from promoting such tokens, GDW integrates a CFG-guided filtering mechanism that identifies syntactically admissible token candidates during decoding. Given a CFG $\mathcal{C}$ specified in extended Backus-Naur form (EBNF), GDW restricts watermark injection to tokens that preserve at least one valid continuation path under $\mathcal{C}$. Example grammar rules are provided in Appendix~\ref{Example Grammar Rules}.

Before decoding, GDW constructs a token-level syntactic mask store from the grammar rules $\mathcal{C}$ and the tokenizer vocabulary. Specifically, lexical specifications associated with CFG terminals are compiled into byte-level finite-state machines (FSMs), and each token $v \in \mathcal{V}$ is decoded into its corresponding byte sequence. By simulating these byte sequences over terminal FSM states, GDW precomputes token-level matching results and stores them in a reusable lookup structure, referred to as the \textsc{MaskStore}. This offline construction efficiently maps grammar-level terminal constraints to vocabulary-level token candidates during decoding.

Given the current prefix code $\mathbf{x}_{1:t-1}$, we define $\mathcal{A}_t \subseteq \mathcal{V}$ as the set of vocabulary tokens that preserve at least one valid continuation path under the CFG $\mathcal{C}$. GDW obtains $\mathcal{A}_t$ through incremental parsing and mask retrieval.

First, an incremental parser takes the current prefix $\mathbf{x}_{1:t-1}$ as input and returns a parse result:

\vspace{-3mm}
\begin{equation}
\mathcal{R}_t=(r_t,\mathcal{P}_t,s_t),
\end{equation}
\vspace{-7mm}

where $r_t$ denotes the prefix-ending remainder fragment that has not yet been fully committed to the CFG parser state. $\mathcal{P}_t$ denotes the set of valid terminal-level continuations derived by the incremental parser from the current CFG parser state. $s_t$ is the interpretation status of $r_t$:

\vspace{-2mm}
\begin{equation}
\begin{aligned}
s_t \in \{&
\textsc{Complete},\, \textsc{Incomplete},\\
&\textsc{MaybeComplete}\}.
\end{aligned}
\end{equation}
\vspace{-4mm}

The parser-returned status $s_t$ determines how the prefix-ending fragment is interpreted during mask retrieval. When $s_t=\textsc{Complete}$, $r_t$ is empty, and the incremental parser determines that the prefix ends at a lexical boundary, so valid candidates are derived from the next terminals in $\mathcal{P}_t$. When $s_t=\textsc{Incomplete}$, the incremental parser identifies $r_t$ as a non-empty fragment that can only be treated as an unfinished terminal. The next token is valid only if appending it to $r_t$ keeps the fragment as a valid terminal prefix or completes it as a CFG terminal.

When $s_t=\textsc{MaybeComplete}$, $r_t$ may correspond to either of the above two paths, and candidates from both paths are retained. For example, in Python, a prefix-ending fragment such as ``return'' may be committed as the keyword \texttt{return}, while it may also be extended by subsequent BPE tokens to form an identifier such as \texttt{return\_value}.

Next, GDW converts these grammar-level constraints into vocabulary-level candidates. The remainder fragment $r_t$ is consumed by the terminal FSMs to obtain the set of reachable FSM states, denoted as $Q(r_t)$. For each reachable FSM state $q \in Q(r_t)$ and each permissible terminal sequence $p \in \mathcal{P}_t$, GDW queries the precomputed \textsc{MaskStore}. The union of retrieved token sets forms the syntactically admissible set:

\begin{equation}
\mathcal{A}_t
=
\bigcup_{q \in Q(r_t)}
\bigcup_{p \in \mathcal{P}_t}
\mathcal{M}(q,p,s_t),
\label{eq:admissible_set}
\end{equation}

where $\mathcal{M}(q,p,s_t)$ denotes the vocabulary-token set retrieved from the precomputed \textsc{MaskStore} for FSM state $q$, terminal sequence $p$, and remainder status $s_t$.

Based on this admissible set $\mathcal{A}_t$, we construct the syntactic validity mask $M_{\mathrm{syn},t} \in \{0,1\}^{|\mathcal{V}|}$. At decoding step $t$,

\begin{equation}
M_{\mathrm{syn},t}^{(v)}
=
\begin{cases}
1, & v \in \mathcal{A}_t,\\
0, & \text{otherwise}.
\end{cases}
\label{eq:syntactic_mask}
\end{equation}

\subsubsection{Structural Role Mask.}

We conduct a detailed analysis of lexical and grammatical conventions of the three programming languages (Python, Java and Go), then partition the vocabulary tokens into two categories: \textbf{syntax-critical tokens}, such as keywords, operators, and delimiters, which constitute the structural skeleton of the program; and \textbf{content-bearing tokens}, including identifiers and literals, which carry specific logic. The complete list of categorized tokens is provided in Appendix \ref{Structural Role Categorization}. To formally incorporate this distinction into our modulation process, we define a structural role mask $M_{\mathrm{role}} \in \{0, 1\}^{|\mathcal{V}|}$ as:
\vspace{-1mm}

\begin{equation}
M_{\mathrm{role}}^{(v)}
=
\begin{cases}
1, & v \in \mathcal{V}_{\mathrm{syntax},}\\
0, & v \in \mathcal{V}_{\mathrm{content}.}
\end{cases}
\label{eq:role_mask}
\end{equation}

\subsubsection{Unified Hierarchical Modulation.}

Based on the three-level masking mechanism, GDW applies hierarchical modulation to determine the valid injection positions and assign role-aware watermark strength. We introduce a \textbf{role function} $r(v)$ to scale the strength based on $M_{\mathrm{role}}^{(v)}$ as: 

\begin{equation}
r(v) = 
\begin{cases}
1, & \text{if } v \in \mathcal{V}_{\mathrm{syntax},}\\
\lambda, & \text{if } v \in \mathcal{V}_{\mathrm{content}.}
\end{cases}
\label{eq:role_function}
\end{equation}

where we assign $\lambda > 1$ to content-bearing tokens to amplify the watermark signal, while applying a more conservative bias to syntax-critical tokens.

Finally, GDW integrates the three masks to compute the watermarked logits from the original logits $l_t$. At decoding step $t$, the watermarked logits $l'_t$ are given by:

\vspace{-2mm}
\begin{equation}
\begin{aligned}
l'_t =\, & l_t
+ M_{\mathrm{wm},t} \odot M_{\mathrm{syn},t} \\
& \odot \left[
M_{\mathrm{role}}
+ \lambda \left(1-M_{\mathrm{role}}\right)
\right]\cdot \delta.
\end{aligned}
\label{eq:final_wm_logits}
\end{equation}

In conclusion, GDW combines the three-level masking mechanism with hierarchical modulation to avoid promoting syntactically invalid tokens and assign role-aware watermark strength, thereby balancing code quality and watermark detectability. The pseudocode of the proposed GDW injection procedure is provided in Algorithm \ref{alg:injection}.

\vspace{-2mm}
\begin{algorithm}[!ht]
\renewcommand{\algorithmicrequire}{\textbf{Input:}}
\renewcommand{\algorithmicensure}{\textbf{Output:}}
\caption{Watermark Injection}
\label{alg:injection}
\begin{algorithmic}[1]
\raggedright

\REQUIRE Prefix tokens $\mathbf{x}_{1:t-1}$, original logits $l_t$, 
green-list ratio $\gamma$, base bias $\delta>0$, 
role function $r(\cdot)$, grammar rules $\mathcal{C}$, 
precomputed \textsc{MaskStore}, syntax-critical vocabulary 
$\mathcal{V}_{\mathrm{syntax}}$, content-bearing vocabulary 
$\mathcal{V}_{\mathrm{content}}$.

\ENSURE Watermarked logits $l'_t$.

\STATE Compute a hash of the preceding token $x_{t-1}$ and use it to pseudorandomly partition the vocabulary $\mathcal{V}$ into green list $G_t$ and red list $R_t$.

\STATE Construct the green-list mask $M_{\mathrm{wm},t}$ by Eq.~\ref{eq:wm_mask}.

\STATE Obtain the incremental parsing result $\mathcal{R}_t$.
\vspace{-2mm}
\[
\begin{aligned}
\mathcal{R}_t &=(r_t,\mathcal{P}_t,s_t),\\
\mathcal{R}_t &\gets
\textsc{IncrementalParser}(\mathcal{C},\mathbf{x}_{1:t-1}).
\end{aligned}
\vspace{-2mm}
\]

\STATE Consume the remainder fragment $r_t$ with the terminal FSMs to obtain the reachable state set $Q(r_t)$.

\STATE Retrieve syntactically admissible token set $\mathcal{A}_t$ from precomputed \textsc{MaskStore} by Eq.~\ref{eq:admissible_set}.

\STATE Construct the syntactic validity mask $M_{\mathrm{syn},t}$ by Eq.~\ref{eq:syntactic_mask}.

\STATE Construct the structural role mask $M_{\mathrm{role}}$ by Eq.~\ref{eq:role_mask}.

\STATE Compute the role-aware scaling factors using the role function in Eq.~\ref{eq:role_function}.

\STATE Integrate the three masks and apply hierarchical modulation to the original logits according to Eq.~\ref{eq:final_wm_logits}.

\RETURN $l'_t$

\end{algorithmic}
\end{algorithm}
\vspace{-4mm}

\subsection{Grammar-Driven Watermark Detection}
\label{Grammar-Driven Watermark Detection}

Given a generated code sequence $\mathbf{x}_{1:T}=(x_1,\dots,x_T)$, we follow the standard practice of discarding the initial token used for seeding. For each subsequent position $t > 1$, GDW recovers the green list $G_t$ using the hash of the preceding token $x_{t-1}$. GDW then weights the contribution of each token $x_t$ according to its structural role, using the same role function $r(\cdot)$ as in the injection phase. We define the weighted detection statistic as:

\vspace{-3mm}
\begin{equation}
z = \frac{ \sum_{t=2}^{T} r(x_t)\left({\textbf{1}}[x_t \in G_t]-\gamma\right)}{
\sqrt{\gamma(1-\gamma)\sum_{t=2}^{T} r(x_t)^2}},
\label{eq:z_score}
\end{equation}

where $\gamma$ is the green-list ratio and $\tau$ is a predefined threshold. A sequence is classified as watermarked if $z >\tau$. The pseudocode of the proposed detection procedure is provided in Algorithm \ref{alg:detection}.

\begin{algorithm}[!ht]
\renewcommand{\algorithmicrequire}{\textbf{Input:}}
\renewcommand{\algorithmicensure}{\textbf{Output:}}
\caption{Watermark Detection}
\label{alg:detection}
\begin{algorithmic}[1]
\REQUIRE Generated code sequence $\mathbf{x}_{1:T}$, green-list ratio $\gamma \in (0, 1)$, role function $r(\cdot)$, detection threshold $\tau$.
\ENSURE Detection result $d \in \{\text{True, False}\}$.

\FOR{$t = 2$ \textbf{to} $T$}
    \STATE Compute a hash of the preceding token $x_{t-1}$.
    
    \STATE Recover the green list $G_t$ using the same pseudorandom partition as in the watermark injection phase.
\ENDFOR

\STATE Calculate the weighted detection statistic \( z \) with Eq.\ref{eq:z_score}.
\IF{$z > \tau$}
    \STATE \textbf{return} True // \textsc{Watermarked Code}.
\ELSE
    \STATE \textbf{return} False // \textsc{Unwatermarked Code}.
\ENDIF
\end{algorithmic}
\end{algorithm}

\section{Experiments}
\subsection{Experimental Settings}

\paragraph{Models.}
We explore the watermark performance using three public LLMs: StarCoder2-3B \citep{lozhkov2024starcoder}, Qwen2.5-Coder-3B \citep{hui2024qwen2}, and Qwen3-4B \citep{Yang2025Qwen3TR}. We additionally report results with Qwen2.5-Coder-7B \citep{hui2024qwen2} in Appendix \ref{Results on Qwen2.5-Coder-7B} to examine whether the observed trend persists at a larger model scale.

\paragraph{Datasets.}
Our method is evaluated across three programming languages: Python, Java, and Go. We use HumanEval \citep{Chen2021EvaluatingLL} and MBPP+ \citep{evalplus,evalperf} for Python, HumanEvalPack \citep{Muennighoff2023OctoPackIT} for Java and Go. We generate $T = 200$ tokens for each sample, using nucleus sampling and beam search.

\begin{table*}[t]
\centering
\small
\setlength{\tabcolsep}{4.5pt} 
\renewcommand{\arraystretch}{1.1} 

\resizebox{\textwidth}{!}{%
\begin{tabular}{lcccccccccccc}
\toprule
\multirow{3}{*}{\raisebox{-4.0ex}{Methods}} 
& \multicolumn{4}{c}{Qwen2.5-Coder-3B} 
& \multicolumn{4}{c}{Qwen3-4B} 
& \multicolumn{4}{c}{StarCoder2-3B} \\
\cmidrule(lr){2-5} 
\cmidrule(lr){6-9} 
\cmidrule(lr){10-13}

& \multicolumn{2}{c}{HumanEval} 
& \multicolumn{2}{c}{MBPP+} 
& \multicolumn{2}{c}{HumanEval} 
& \multicolumn{2}{c}{MBPP+} 
& \multicolumn{2}{c}{HumanEval} 
& \multicolumn{2}{c}{MBPP+} \\
\cmidrule(lr){2-3} 
\cmidrule(lr){4-5} 
\cmidrule(lr){6-7} 
\cmidrule(lr){8-9} 
\cmidrule(lr){10-11} 
\cmidrule(lr){12-13}

& \raisebox{0.9ex}{\shortstack[c]{Sampling}} & \shortstack{Beam\\Search} 
& \raisebox{0.9ex}{\shortstack[c]{Sampling}} & \shortstack{Beam\\Search} 
& \raisebox{0.9ex}{\shortstack[c]{Sampling}} & \shortstack{Beam\\Search} 
& \raisebox{0.9ex}{\shortstack[c]{Sampling}} & \shortstack{Beam\\Search} 
& \raisebox{0.9ex}{\shortstack[c]{Sampling}} & \shortstack{Beam\\Search} 
& \raisebox{0.9ex}{\shortstack[c]{Sampling}} & \shortstack{Beam\\Search}  \\
\midrule

KGW   
& 0.745 & 0.832 
& 0.735 & 0.857 
& 0.714 & 0.815 
& 0.784 & 0.840 
& 0.691 & 0.848 
& 0.750 & 0.837 \\

SWEET 
& 0.784 & 0.742 
& 0.757 & 0.768 
& 0.766 & 0.756 
& 0.719 & 0.774 
& 0.722 & 0.743 
& 0.761 & 0.784 \\

EWD   
& 0.771 & 0.774 
& 0.763 & 0.789 
& 0.796 & 0.831 
& 0.797 & 0.852 
& 0.712 & 0.795 
& 0.743 & 0.820 \\

STONE 
& 0.666 & 0.677 
& 0.670 & 0.670 
& 0.666 & 0.671 
& 0.669 & 0.683 
& 0.681 & 0.686 
& 0.668 & 0.683 \\

CodeIP 
& 0.805 & 0.730 
& 0.767 & 0.768 
& 0.722 & 0.802 
& 0.809 & 0.796 
& 0.725 & 0.760 
& 0.766 & 0.739 \\

SynthID-Text 
& 0.702 & 0.692 
& 0.676 & 0.668 
& 0.705 & 0.701 
& 0.744 & 0.700 
& 0.700 & 0.679 
& 0.732 & 0.754 \\

GDW   
& \textbf{0.811} & \textbf{0.847} 
& \textbf{0.787} & \textbf{0.876} 
& \textbf{0.830} & \textbf{0.847} 
& \textbf{0.828} & \textbf{0.875} 
& \textbf{0.743} & \textbf{0.862} 
& \textbf{0.833} & \textbf{0.855} \\

\bottomrule
\end{tabular}%
}

\caption{Area Under the Trade-off Curve (AUTC) of different watermarking methods on HumanEval and MBPP+. The best trade-off performance in each column is highlighted in bold.}
\label{tab:autc_python}
\end{table*}

\begin{figure*}[t]
\centering
\setlength{\tabcolsep}{1pt}

\begin{subfigure}[t]{0.16\textwidth}
    \centering
    \includegraphics[width=\linewidth]{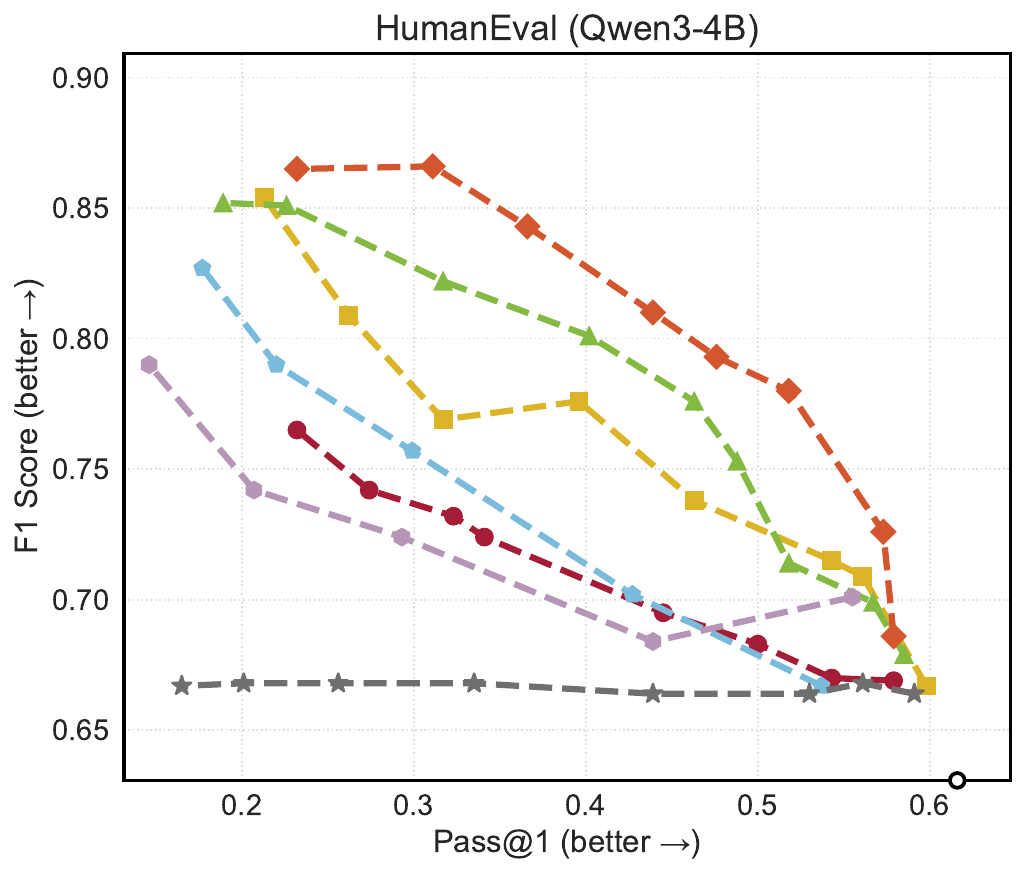}
\end{subfigure}
\hfill
\begin{subfigure}[t]{0.16\textwidth}
    \centering
    \includegraphics[width=\linewidth]{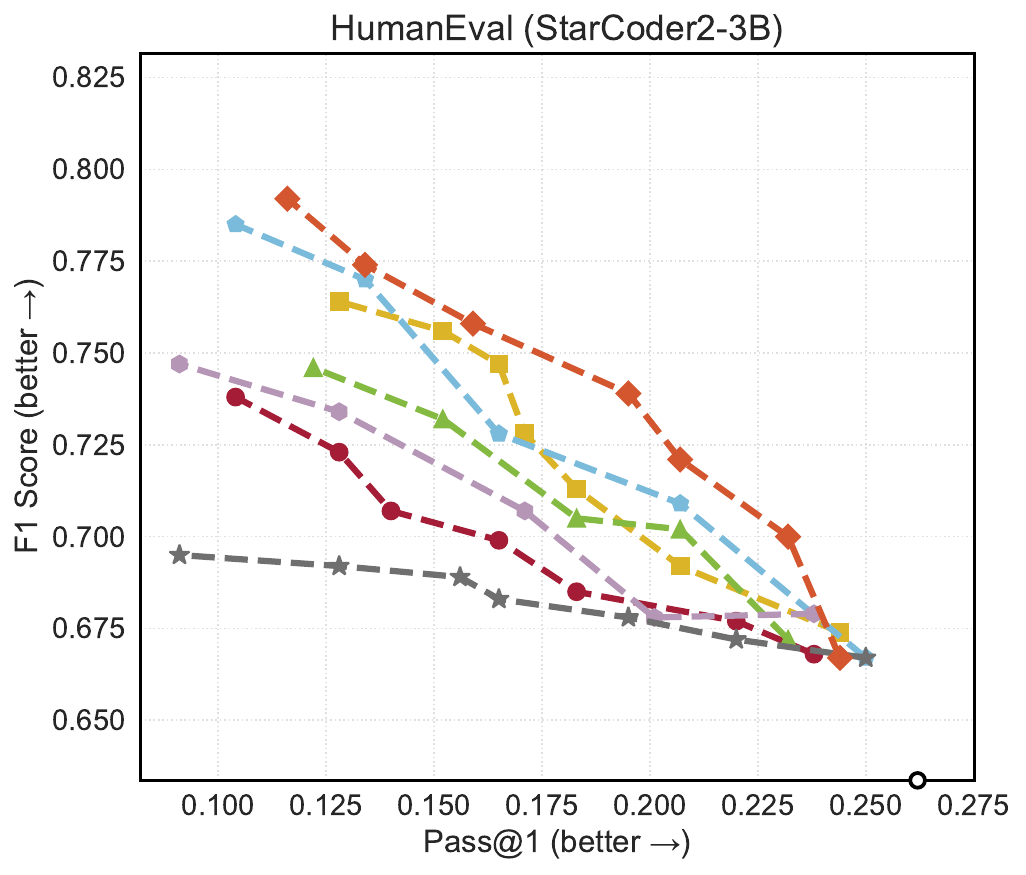}
\end{subfigure}
\hfill
\begin{subfigure}[t]{0.16\textwidth}
    \centering
    \includegraphics[width=\linewidth]{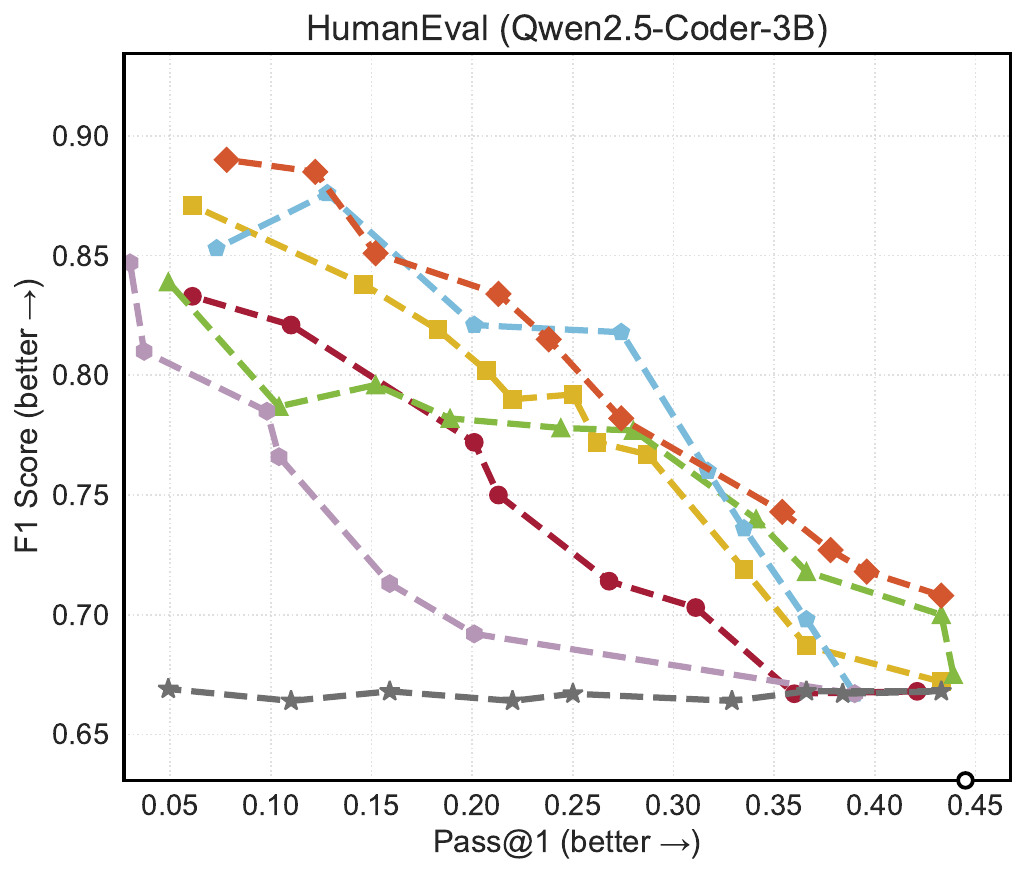}
\end{subfigure}
\hfill
\begin{subfigure}[t]{0.16\textwidth}
    \centering
    \includegraphics[width=\linewidth]{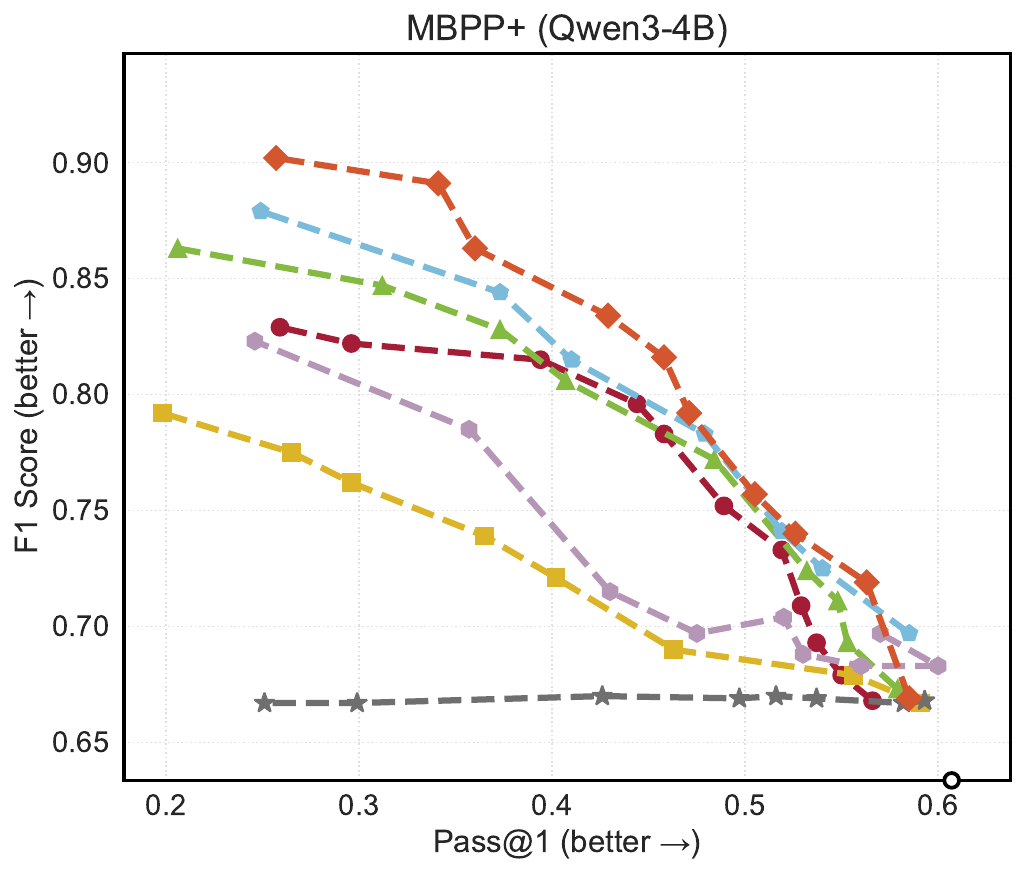}
\end{subfigure}
\hfill
\begin{subfigure}[t]{0.16\textwidth}
    \centering
    \includegraphics[width=\linewidth]{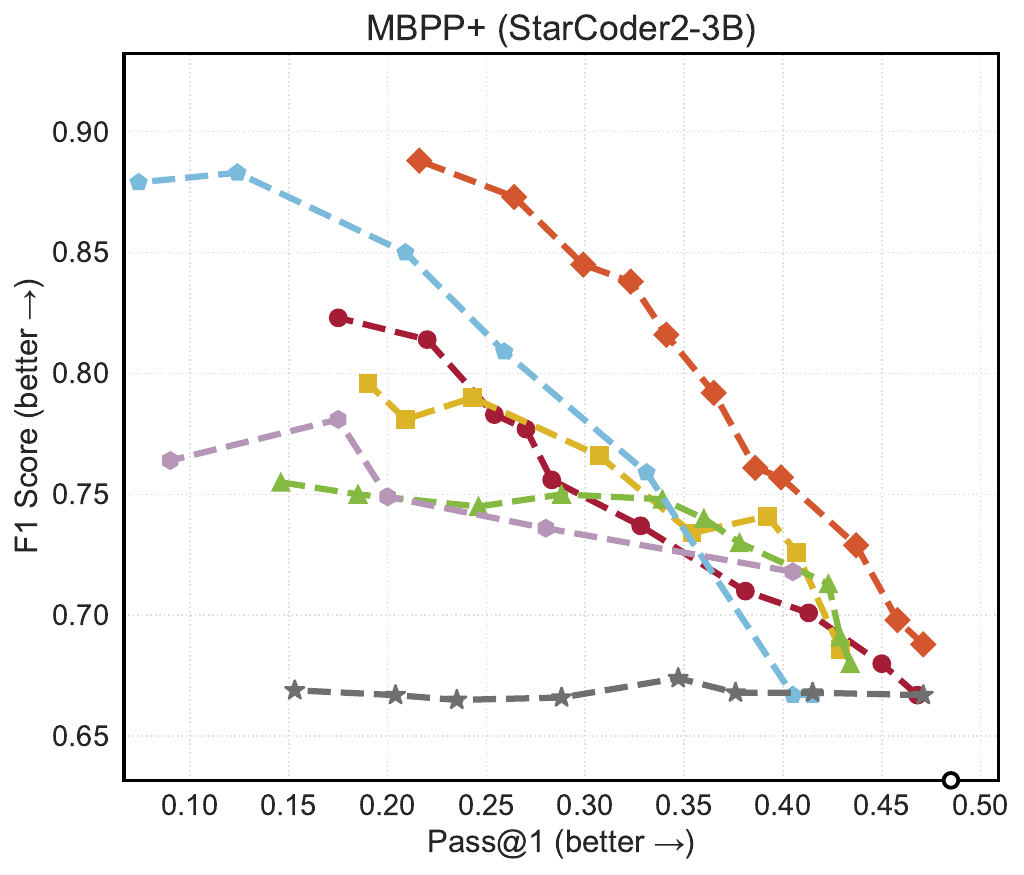}
\end{subfigure}
\hfill
\begin{subfigure}[t]{0.16\textwidth}
    \centering
    \includegraphics[width=\linewidth]{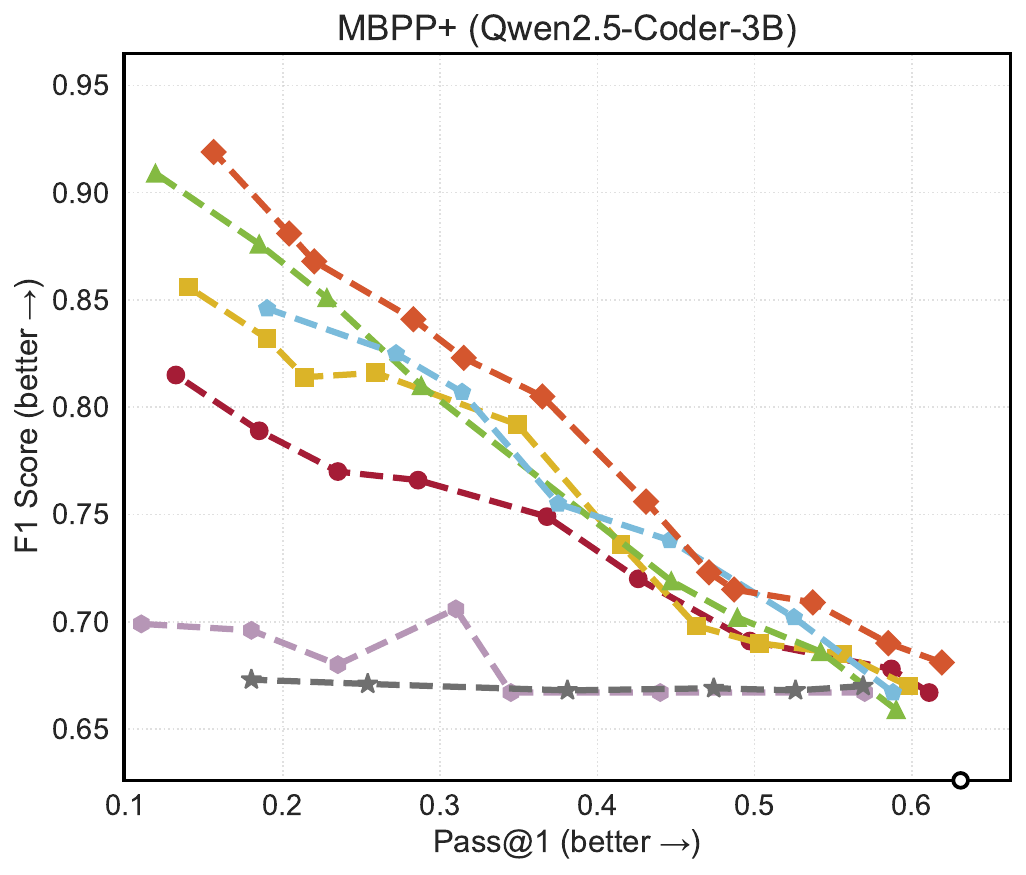}
\end{subfigure}

\begin{subfigure}[t]{0.16\textwidth}
    \centering
    \includegraphics[width=\linewidth]{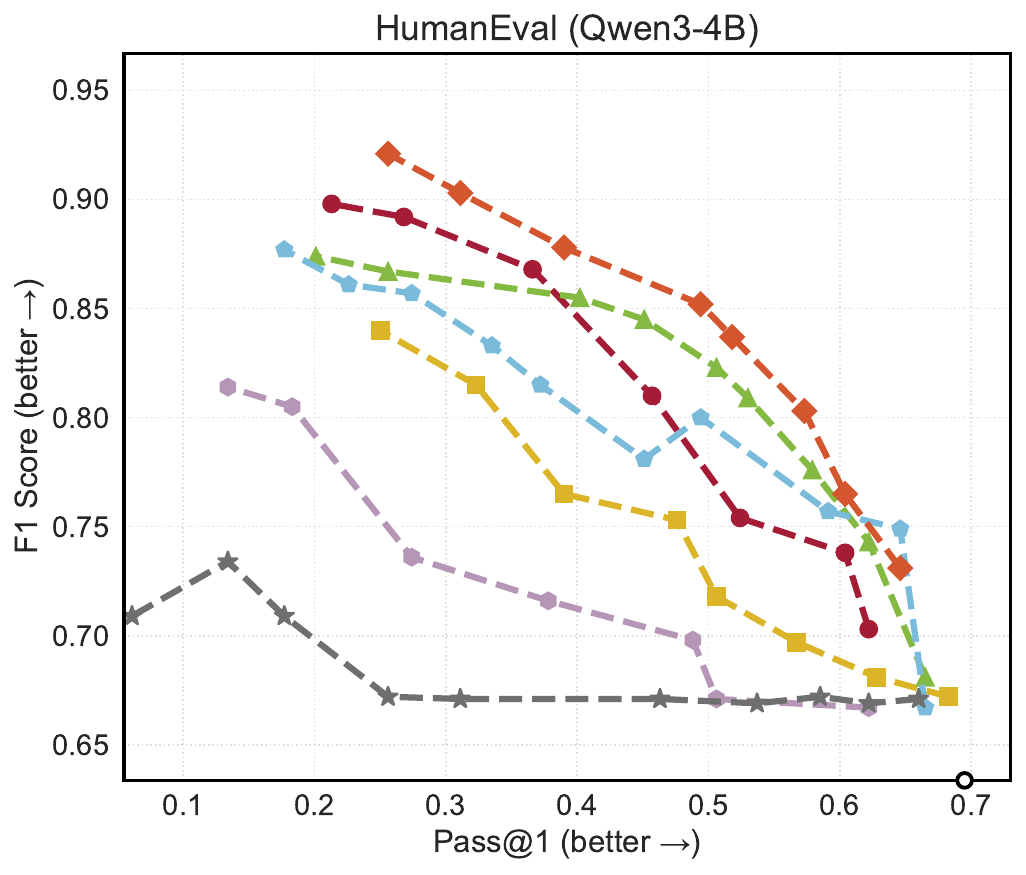}
\end{subfigure}
\hfill
\begin{subfigure}[t]{0.16\textwidth}
    \centering
    \includegraphics[width=\linewidth]{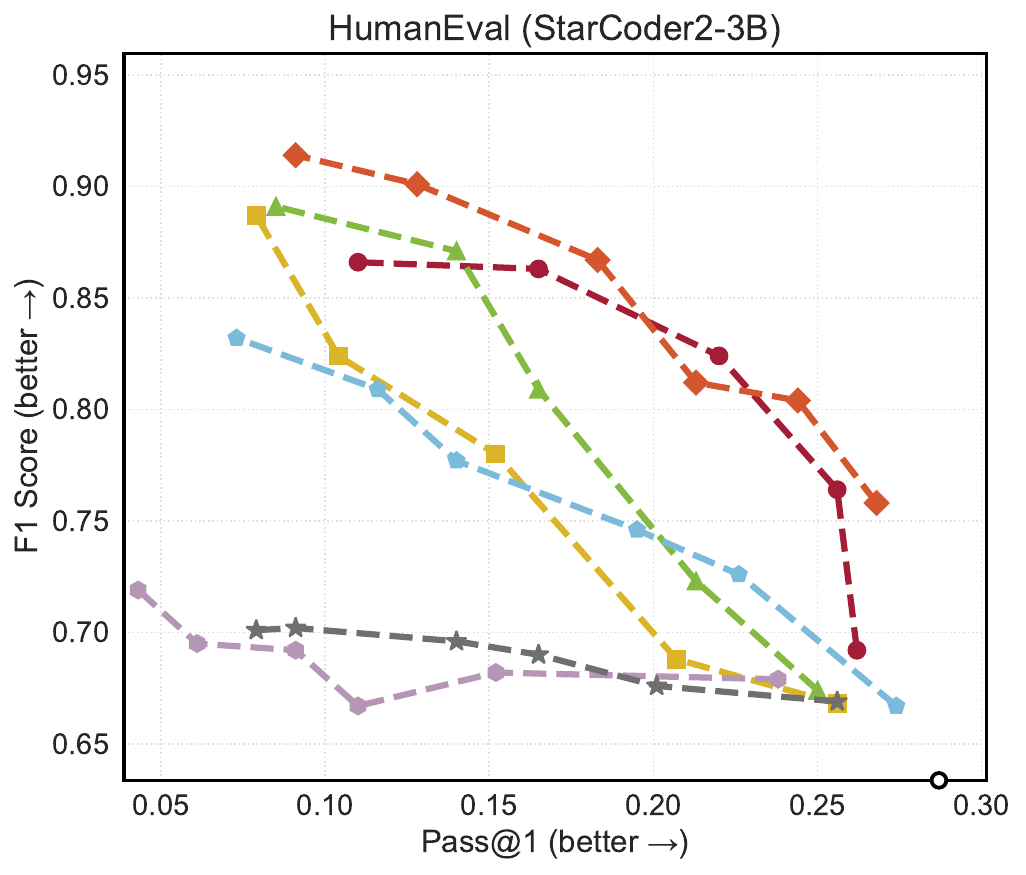}
\end{subfigure}
\hfill
\begin{subfigure}[t]{0.16\textwidth}
    \centering
    \includegraphics[width=\linewidth]{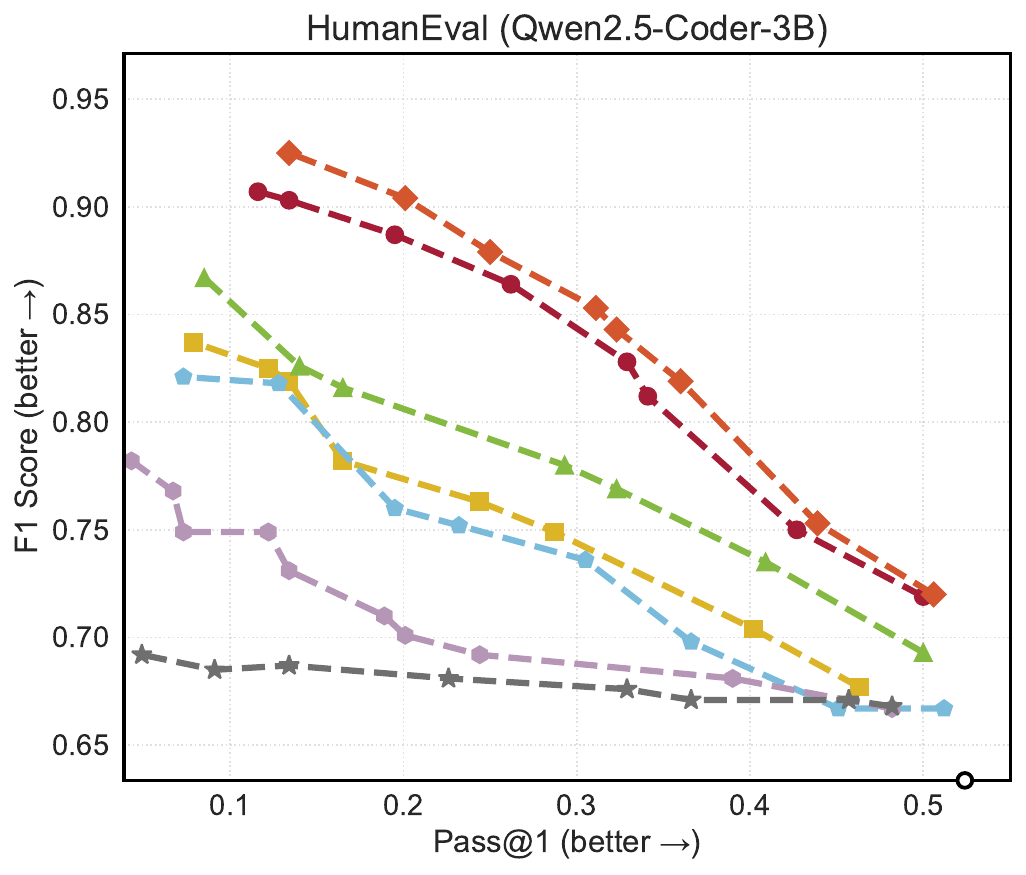}
\end{subfigure}
\hfill
\begin{subfigure}[t]{0.16\textwidth}
    \centering
    \includegraphics[width=\linewidth]{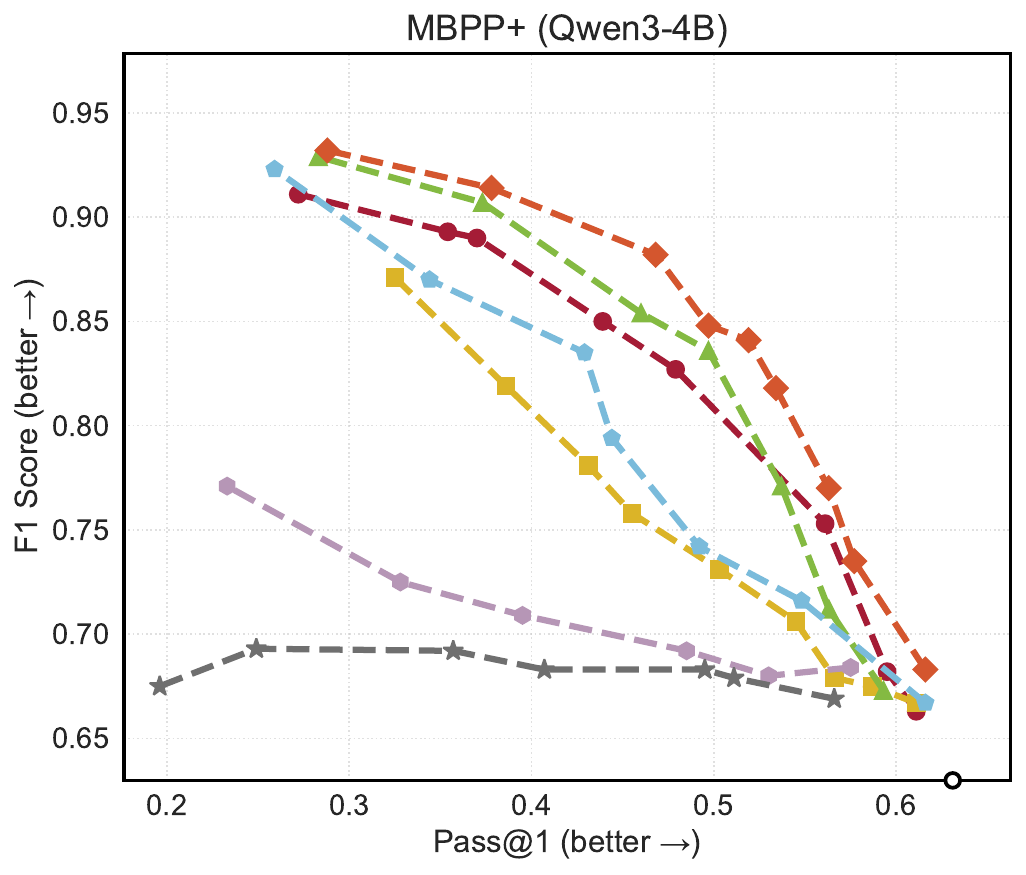}
\end{subfigure}
\hfill
\begin{subfigure}[t]{0.16\textwidth}
    \centering
    \includegraphics[width=\linewidth]{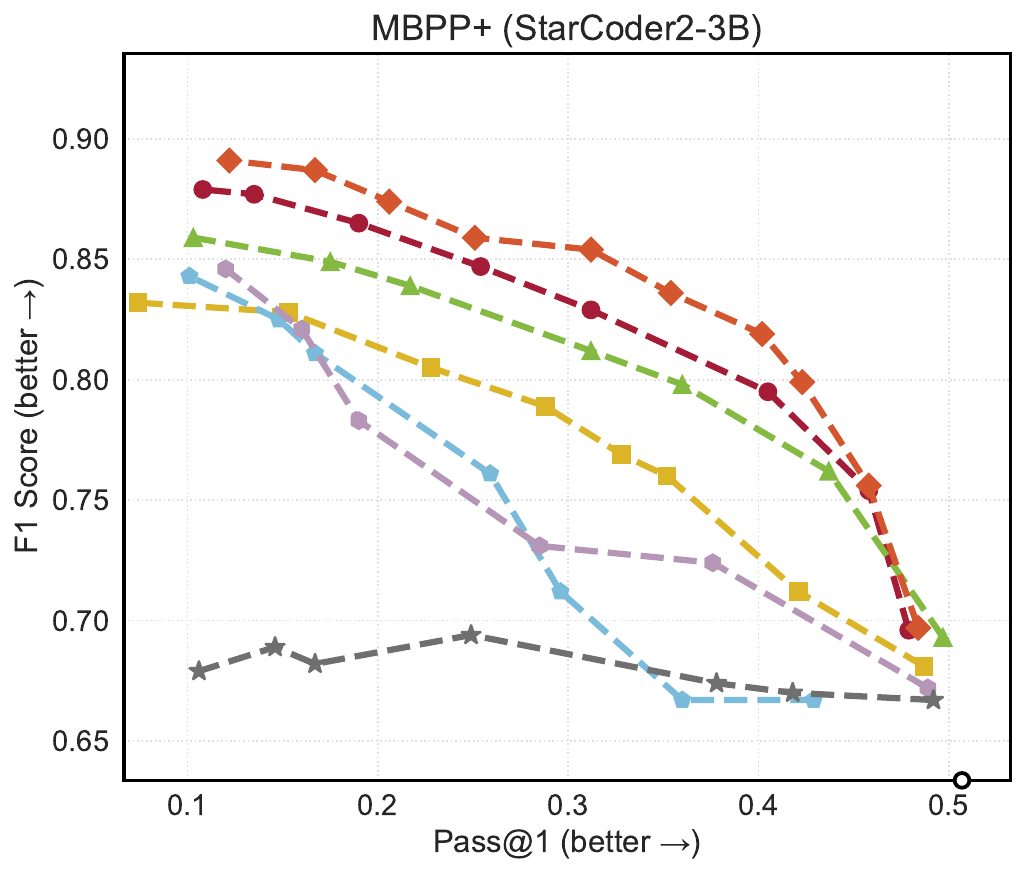}
\end{subfigure}
\hfill
\begin{subfigure}[t]{0.16\textwidth}
    \centering
    \includegraphics[width=\linewidth]{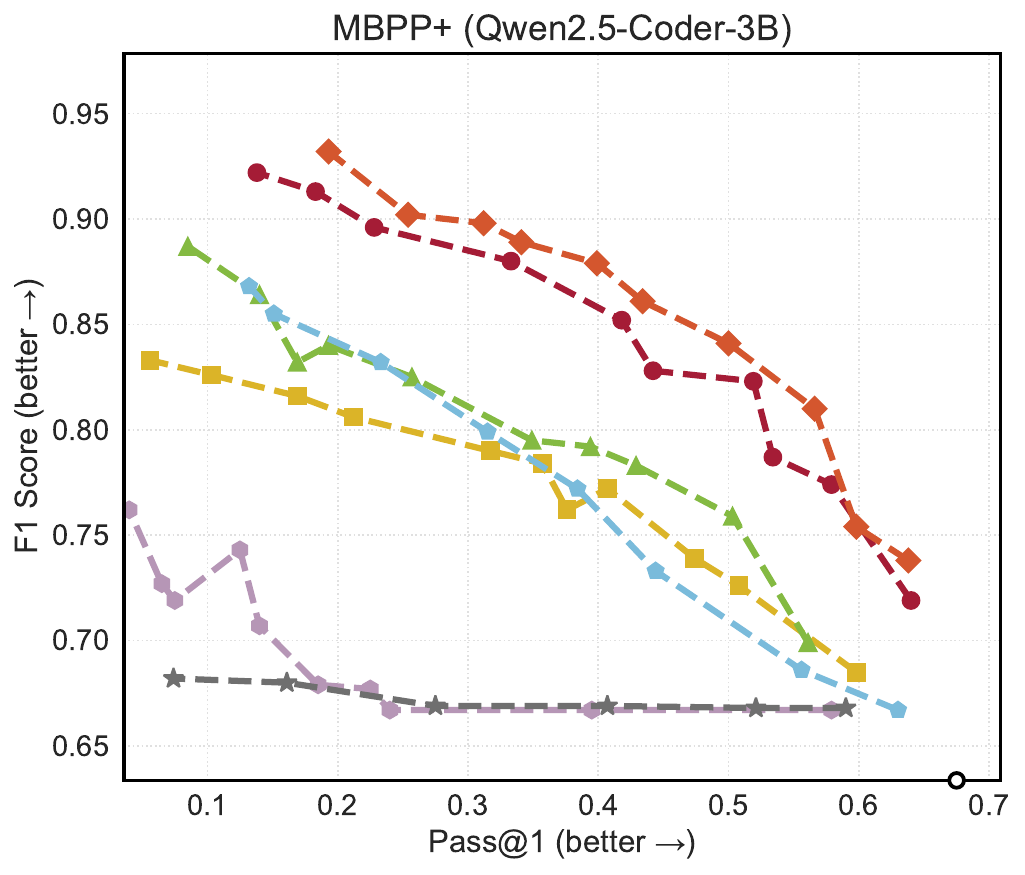}
\end{subfigure}

\begin{center}
\scriptsize
\setlength{\tabcolsep}{8pt}
\newcommand{\legendwithline}[3]{
\textcolor[HTML]{#1}{
\tikz[baseline=-0.6ex]{
\draw[line width=1pt] (0,0) -- (2.5em,0);
\node at (1.25em,0) {#2};
}}
\hspace{0.2em} #3
}
\begin{tabular}{ccccccc}
\legendwithline{A51C36}{$\bullet$}{KGW}
&
\legendwithline{DBB428}{$\blacksquare$}{SWEET}
&
\legendwithline{84BA42}{$\blacktriangle$}{EWD}
&
\legendwithline{6F6F6F}{$\bigstar$}{STONE}
&
\legendwithline{7ABBDB}{
\tikz\filldraw (90:0.7ex)
-- (18:0.7ex)
-- (-54:0.7ex)
-- (-126:0.7ex)
-- (162:0.7ex)
-- cycle;
}{CodeIP}
&
\legendwithline{B696B6}{
\tikz\filldraw
(0:0.7ex)
-- (60:0.7ex)
-- (120:0.7ex)
-- (180:0.7ex)
-- (240:0.7ex)
-- (300:0.7ex)
-- cycle;
}{SynthID-Text}
&
\legendwithline{D4562E}{$\blacklozenge$}{GDW (Ours)}
\end{tabular}
\end{center}
\caption{Pass@1-F1 quality-detectability trade-off curves of watermarked code generation under sampling (top row) and beam search (bottom row). The hollow marker represents the unwatermarked pass@1 performance as a quality reference. Curves closer to the upper-right region indicate a more favorable trade-off.}
\label{fig:Python_tradeoff}
\end{figure*}

\paragraph{Metrics.}
In code generation tasks, executability and functional correctness are the primary concerns. Therefore, we adopt pass@1 \citep{Chen2021EvaluatingLL} as the main metric for code quality. For watermark detectability, we report the F1 score at a fixed false positive rate of 1\%.

Since watermark strength affects both code quality and watermark detectability, a single operating point cannot fully characterize their trade-off. We therefore vary the watermark strength to construct a Pass@1-F1 trade-off curve. To quantify this curve, we introduce the Area Under the Trade-off Curve (AUTC), defined as the normalized area under the Pass@1-F1 trade-off curve over the common quality interval $[Q_{\text{start}}, Q_{\text{end}}]$. The boundaries of this interval are determined as follows:

\vspace{-6mm}
\begin{equation}
Q_{\mathrm{start}}
=
\max_m \min_i q_{m,i},
\quad
Q_{\mathrm{end}}
=
\min_m \max_i q_{m,i},
\label{eq:common_interval}
\end{equation} 
where $q_{m,i}$ denotes the pass@1 value of method $m$ at operating point $i$.

Formally, the Area Under the Trade-off Curve (AUTC) of method $m$ is computed as:

\vspace{-6mm}
\begin{equation}
\mathrm{AUTC}_m
=
\frac{1}{Q_{\mathrm{end}}-Q_{\mathrm{start}}}
\int_{Q_{\mathrm{start}}}^{Q_{\mathrm{end}}}
\widetilde{\mathrm{F1}}_m(q)\,\mathrm{d}q,
\label{eq:autc}
\end{equation}

where $\widetilde{\mathrm{F1}}_m(q)$ denotes the interpolated detection F1 at the pass@1 quality $q$. AUTC values computed over the same common quality interval are directly comparable, with higher AUTC indicating a more favorable quality-detectability trade-off. The common quality interval for each experiment is reported in Appendix \ref{Common Quality Intervals Used for AUTC Computation}. We also provide further discussion of the AUTC metric in Appendix \ref{Discussion on AUTC Metric}.

\begin{table*}[t]
\centering
\small
\setlength{\tabcolsep}{4.8pt}
\renewcommand{\arraystretch}{1.1}
\begin{tabular}{lcccccccc}
\toprule
\multirow{3}{*}{\raisebox{-2.0ex}{Methods}} 
& \multicolumn{4}{c}{Qwen2.5-Coder-3B} 
& \multicolumn{4}{c}{StarCoder2-3B} \\
\cmidrule(lr){2-5} \cmidrule(lr){6-9}
& \multicolumn{2}{c}{HumanEvalPack-Java} 
& \multicolumn{2}{c}{HumanEvalPack-Go} 
& \multicolumn{2}{c}{HumanEvalPack-Java} 
& \multicolumn{2}{c}{HumanEvalPack-Go} \\
\cmidrule(lr){2-3} \cmidrule(lr){4-5} 
\cmidrule(lr){6-7} \cmidrule(lr){8-9}
& Sampling & BeamSearch 
& Sampling & BeamSearch 
& Sampling & BeamSearch 
& Sampling & BeamSearch \\
\midrule
KGW   & 0.768 & 0.870 & 0.911 & 0.910 & 0.797 & 0.857 & 0.725 & 0.861 \\
SWEET & 0.806 & 0.781 & 0.842 & 0.836 & 0.813 & 0.828 & 0.779 & 0.721 \\
EWD   & 0.865 & 0.854 & 0.881 & 0.879 & 0.834 & 0.859 & 0.770 & 0.748 \\
STONE & 0.675 & 0.781 & ---   & ---   & 0.679 & 0.714 & ---   & ---   \\
CodeIP & 0.687 & 0.750 & 0.726 & 0.768 & 0.667 & 0.682 & 0.668 & 0.723 \\
SynthID-Text & 0.721 & 0.813 & 0.725 & 0.729 & 0.783 & 0.702 & 0.689 & 0.680 \\
GDW
& \textbf{0.887} & \textbf{0.910} 
& \textbf{0.951} & \textbf{0.949} 
& \textbf{0.871} & \textbf{0.900} 
& \textbf{0.831} & \textbf{0.881} \\
\bottomrule
\end{tabular}
\caption{Area Under the Trade-off Curve (AUTC) of different watermarking methods on HumanEvalPack.}
\label{tab:autc_humanevalpack}
\vspace{-2mm}
\end{table*}

\begin{table*}[t]
\centering
\small
\begin{tabular*}{\textwidth}{
@{\extracolsep{\fill}}
lcccccccc
@{}
}
\toprule
\raisebox{+0.8\height}{Method}
& \shortstack{No\\Attack}
& \shortstack{Rename\\Variables}
& \shortstack{Add Dead\\Code}
& \shortstack{Insert Print\\Statements}
& \shortstack{Unroll While\\Loops}
& \shortstack{Wrap with\\Try--Catch}
& \shortstack{Replace\\True/False}
& \shortstack{Remove\\Comments} \\
\midrule
KGW
& 0.852 & 0.834 & 0.716 & 0.824
& 0.693 & 0.701 & 0.710 & 0.720 \\

SWEET
& 0.754 & 0.687 & 0.702 & 0.703
& 0.685 & 0.703 & 0.706 & 0.703 \\

EWD
& 0.786 & 0.771 & 0.676 & 0.764
& 0.682 & 0.680 & 0.681 & 0.680 \\

STONE
& 0.679 & 0.674 & 0.671 & 0.675
& 0.671 & 0.670 & 0.672 & 0.674 \\

CodeIP
& 0.743 & 0.701 & 0.667 & 0.701
& 0.671 & 0.673 & 0.674 & 0.667 \\

SynthID-Text
& 0.696 & 0.687 & 0.680 & 0.688
& 0.674 & 0.676 & 0.680 & 0.679 \\

GDW
& \textbf{0.869} & \textbf{0.837} & \textbf{0.727} & \textbf{0.839}
& \textbf{0.704} & \textbf{0.724} & \textbf{0.729} & \textbf{0.736} \\
\bottomrule
\end{tabular*}


\caption{Area Under the Trade-off Curve (AUTC) of different watermarking methods under various semantics-preserving attacks on HumanEval using Qwen2.5-Coder-3B with beam search decoding. All columns are computed over the same quality interval [0.134, 0.396].}

\label{tab:robustness}
\vspace{-3mm}
\end{table*}

\paragraph{Baselines.}
We compare GDW with six watermarking methods, including the pioneering logits-based watermark KGW \citep{kirchenbauer2023watermark}, the entropy-based watermark SWEET \citep{lee2024wrote} and EWD \citep{lu2024entropy}, the code watermark STONE \citep{kim2026marking} and CodeIP \citep{guan2024codeip}, as well as the non-red-green watermark SynthID-Text \citep{dathathri2024scalable}.

\paragraph{Hyperparameters.}
For red-green watermarking methods, we set the green-list ratio $\gamma=0.5$ and evaluate F1 at a false positive rate of 1\%, following~\citet{kirchenbauer2023watermark}. To construct the quality-detectability curve, we sweep the watermark-strength $\delta \in (0, 5.0]$. For SynthID-Text, we sweep the number of watermark keys in $[1,30]$. We set $\lambda=2.0$ for GDW. More experimental details are provided in Appendix~\ref{Details of Experimental Settings}.

\subsection{Overall Trade-off Performance}
We evaluate GDW and six baselines across benchmarks covering Python, Java, and Go. Tables~\ref{tab:autc_python} and~\ref{tab:autc_humanevalpack} respectively report the AUTC results on Python benchmarks and multilingual HumanEvalPack benchmarks. GDW achieves the highest AUTC across all settings. Figure~\ref{fig:Python_tradeoff} and~\ref{fig:HumanEvalPack_Trade-off} further visualize the trade-off curves, with unwatermarked pass@1 results provided as quality references. GDW achieves a more favorable trade-off between code quality and watermark detectability than the baselines.

\begin{figure}[t]
    \centering

    \begin{subfigure}[t]{0.48\columnwidth}
        \centering
        \includegraphics[width=\linewidth]{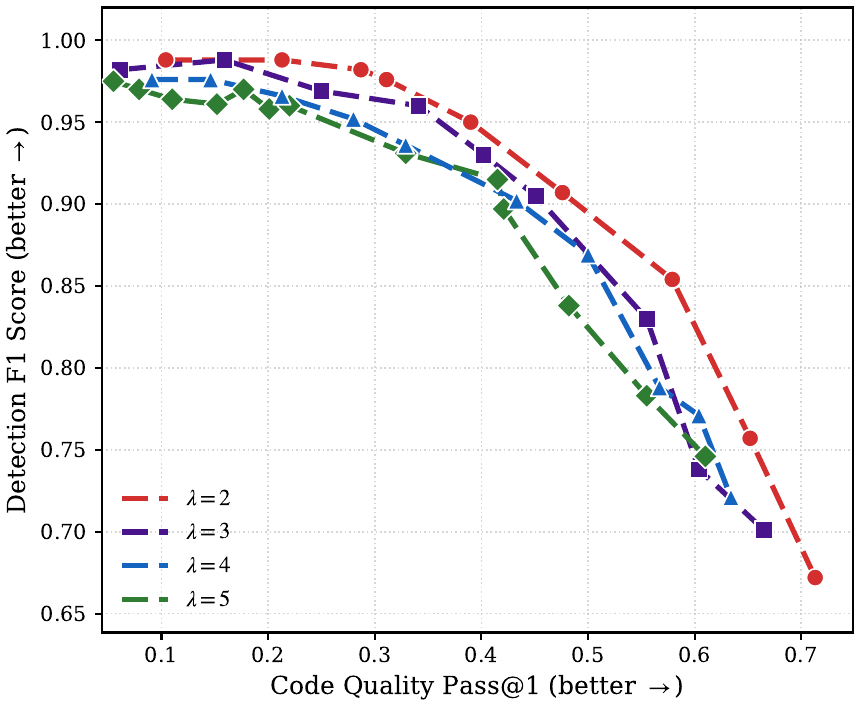}
        \caption{}
        \label{fig:Lambda_Comp_a}
    \end{subfigure}
    \hfill
    \begin{subfigure}[t]{0.48\columnwidth}
        \centering
        \includegraphics[width=\linewidth]{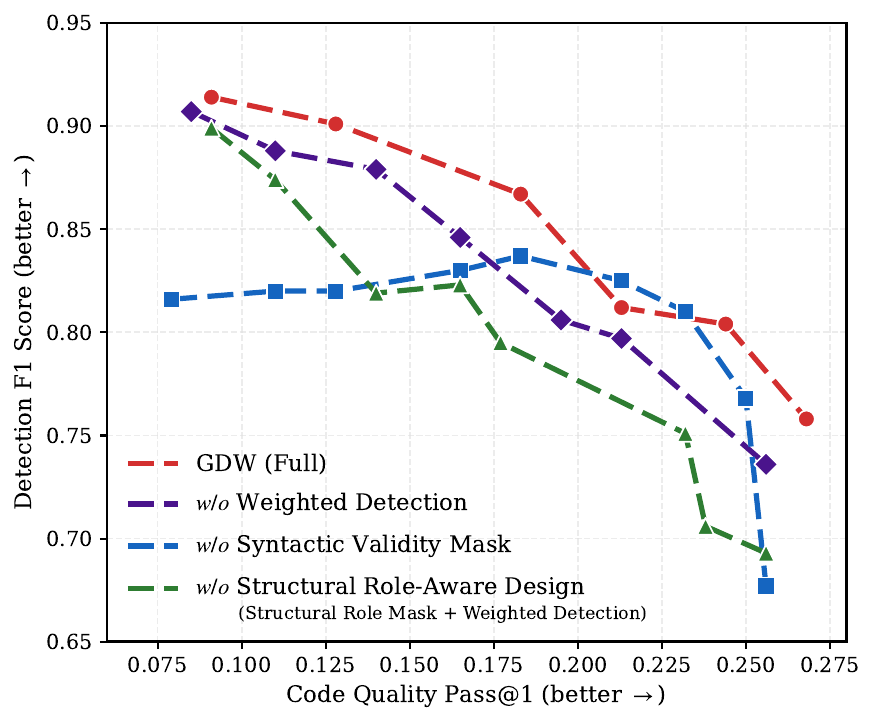}
        \caption{}
        \label{fig:Ablation_Study_b}
    \end{subfigure}

    \caption{
    Subfigure (a) illustrates the trade-off curves of GDW for different $\lambda$ values. 
    Subfigure (b) shows the trade-off curves of GDW and its ablated variants on StarCoder2-3B using HumanEval with beam search.
    }
    \label{fig:combined}
    \vspace{-3mm}
\end{figure}

\subsection{Effect of the role scaling factor \textbf{\(\lambda\)}}

We study the role scaling factor $\lambda$, which controls the additional watermark bias assigned to content-bearing tokens. Figure~\ref{fig:Lambda_Comp_a} and Table~\ref{tab:lambda_sensitivity_java} report the results for $\lambda \in \{2,3,4,5\}$. Across all settings, $\lambda=2$ yields the highest AUTC score. Increasing $\lambda$ monotonically decreases AUTC from 0.942 to 0.897, indicating that moderately increasing the watermark bias on content-bearing tokens is beneficial, whereas overly aggressive modulation degrades the overall balance between detectability and code quality. We therefore select $\lambda=2$ as the default value in the experiments.

\begin{table}[t]
\centering
\small
\setlength{\tabcolsep}{4pt}
\renewcommand{\arraystretch}{1.15}
\begin{tabular*}{\columnwidth}{
@{\extracolsep{\fill}}
lcccc
@{}
}
\toprule
\textbf{$\lambda$}
& \textbf{$\lambda=2$}
& \textbf{$\lambda=3$}
& \textbf{$\lambda=4$}
& \textbf{$\lambda=5$} \\
\midrule
AUTC & \textbf{0.942} & 0.922 & 0.910 & 0.897 \\
\bottomrule
\end{tabular*}
\caption{AUTC under different values of the role scaling factor $\lambda$ on HumanEvalPack-Java using Qwen2.5-Coder-3B with beam search decoding.}
\label{tab:lambda_sensitivity_java}
\end{table}

\subsection{Ablation Study}
We conduct ablations to study the contribution of each GDW component. As shown in Table~\ref{tab:ablation_autc}, removing the Weighted Detection module decreases AUTC from 0.860 to 0.832, demonstrating the benefit of aligning the detection statistic with the hierarchical injection strategy. Removing the Syntactic Validity Mask decreases AUTC to 0.817, showing that grammar-driven filtering contributes to maintaining a favorable quality-detectability trade-off. Removing the Structural Role-Aware Design results in the largest drop, from 0.860 to 0.800, highlighting the importance of incorporating structural roles into watermark injection and detection.

\begin{table}[t]
\centering
\small
\setlength{\tabcolsep}{4pt}
\renewcommand{\arraystretch}{1.15}
\begin{tabular*}{\columnwidth}{
@{\extracolsep{\fill}}
lc
@{}
}
\toprule
Variant & AUTC $\uparrow$ \\
\midrule
GDW (Full)                  & \textbf{0.860} \\
w/o Weighted Detection      & 0.832 \\
w/o Syntactic Validity Mask & 0.817 \\
w/o Structural Role-Aware Design       & 0.800 \\
\bottomrule
\end{tabular*}
\caption{Ablation study of GDW on HumanEval using StarCoder2-3B with beam search decoding. Note that the Role-Aware Design variant removes both the Structural Role Mask and the aligned Weighted Detection.}
\label{tab:ablation_autc}
\end{table}

\begin{table}[t]
\centering
\small
\setlength{\tabcolsep}{7pt}
\renewcommand{\arraystretch}{1.2}
\begin{tabular}{lcc}
\toprule
Method & \shortstack{Latency (ms/token) $\downarrow$} & Slowdown $\downarrow$ \\
\midrule
No Watermark & 18.09 & 1.000$\times$\\
KGW & 18.91 & 1.045$\times$\\
SWEET & 18.97 & 1.049$\times$\\
EWD & 18.88 & 1.044$\times$\\
STONE & 18.96 & 1.048$\times$\\
SynthID-Text & 21.98 & 1.215$\times$\\
CodeIP & 19.24 & 1.064$\times$\\
GDW & 20.67 & 1.143$\times$\\
\bottomrule
\end{tabular}
\caption{Latency comparison on MBPP+ using Qwen2.5-Coder-3B with nucleus sampling. Slowdown is the latency ratio relative to no-watermark inference.}
\label{tab:inference_efficiency}
\end{table}

\subsection{Robustness Analysis}
We evaluate GDW under seven semantics-preserving attacks following \citep{suresh2024watermarking}. As shown in Table~\ref{tab:robustness}, GDW achieves the highest AUTC under all evaluated attacks, indicating that its trade-off advantage remains stable. Compared with the no-attack setting, GDW's AUTC decreases from 0.869 to 0.839 under print-statement insertion and to 0.837 under variable renaming. In contrast, attacks involving larger structural changes, including dead-code insertion, loop unrolling, and try-catch wrapping, cause greater degradation across methods. Nevertheless, GDW maintains the best quality-detectability trade-off under these more structure-altering transformations.

\subsection{Inference Efficiency}
We measure inference latency on Qwen2.5-Coder-3B with the MBPP+ dataset under nucleus sampling, using a single NVIDIA RTX 5090 GPU. As shown in Table~\ref{tab:inference_efficiency}, GDW increases latency from 18.09 ms/token without watermarking to 20.67 ms/token, corresponding to a 1.143$\times$ slowdown. The incremental parser accounts for 1.84 ms/token on average and is the main source of the additional overhead. Compared with the baselines, GDW incurs slightly higher latency than lightweight red-green methods but remains faster than SynthID-Text. This suggests that syntactic validity masking introduces a measurable but bounded overhead, which could be further reduced with more efficient incremental parsers or parallelized mask retrieval.

\section{Conclusion}
In this paper, we propose GDW to watermark the LLMs for code generation. By decomposing watermark injection into \emph{where} to inject and \emph{how strongly} to bias tokens, GDW avoids promoting syntactically invalid tokens and applies structural role-aware modulation to balance code quality and detectability. Aligned with the injection strategy, we design a role-aware detection statistic. For evaluation, we introduce AUTC as a unified metric to quantify the quality-detectability trade-off. Comprehensive experiments show that GDW establishes a stronger quality-detectability trade-off frontier than existing methods and remains robust under various semantics-preserving attacks. Our findings contribute to more reliable identification of LLM-generated code in the era of LLM-assisted software development.

\section*{Limitations}
In this work, we proposed GDW, a grammar-driven watermarking approach that achieves a superior trade-off between code quality and detectability. A potential limitation is that while GDW consistently outperforms baselines across diverse models and benchmarks, it introduces a slight computational overhead during the inference stage due to the necessity of incremental grammar parsing. Future optimization could involve the use of more efficient, specialized incremental parsers or parallelization techniques to further minimize this overhead.

\section*{Ethical Considerations}
We use open-source artifacts in accordance with their licenses and intended research use. The data resources used in this study are publicly available.


\bibliography{custom}

\begin{thebibliography}{41}
\providecommand{\natexlab}[1]{#1}

\bibitem[{Anthropic(2025)}]{ClaudeCode}
Anthropic. 2025.
\newblock Claude code.
\newblock \url{https://claude.com/product/claude-code}.
\newblock [Accessed 22-05-2026].

\bibitem[{Anysphere(2023)}]{Cursor}
Anysphere. 2023.
\newblock Cursor.
\newblock \url{https://www.cursor.com/}.
\newblock [Accessed 22-05-2026].

\bibitem[{Balachandran et~al.(2014)Balachandran, Keong, and
  Emmanuel}]{balachandran2014function}
Vivek Balachandran, Ng~Wee Keong, and Sabu Emmanuel. 2014.
\newblock Function level control flow obfuscation for software security.
\newblock In \emph{2014 Eighth International Conference on Complex, Intelligent
  and Software Intensive Systems}, pages 133--140. IEEE.

\bibitem[{Brown et~al.(1990)Brown, Cocke, Della~Pietra, Della~Pietra, Jelinek,
  Lafferty, Mercer, and Roossin}]{brown1990statistical}
Peter~F Brown, John Cocke, Stephen~A Della~Pietra, Vincent~J Della~Pietra,
  Frederick Jelinek, John Lafferty, Robert~L Mercer, and Paul~S Roossin. 1990.
\newblock A statistical approach to machine translation.
\newblock \emph{Computational linguistics}, 16(2):79--85.

\bibitem[{Chen et~al.(2017)Chen, Li, Wen, Chen, and Yan}]{chen2017software}
Jianping Chen, Kui Li, Wanzhi Wen, Weixu Chen, and Chenxue Yan. 2017.
\newblock Software watermarking for java program based on method name encoding.
\newblock In \emph{International Conference on Advanced Intelligent Systems and
  Informatics}, pages 865--874. Springer.

\bibitem[{Chen et~al.(2021)Chen, Tworek, Jun, Yuan, Pond{\'e}, Kaplan, Edwards,
  Burda, Joseph, Brockman, Ray, Puri, Krueger, Petrov, Khlaaf, Sastry, Mishkin,
  Chan, Gray, Ryder, Pavlov, Power, Kaiser, Bavarian, Winter, Tillet, Such,
  Cummings, Plappert, Chantzis, Barnes, Herbert-Voss, Guss, Nichol, Babuschkin,
  Balaji, Jain, Carr, Leike, Achiam, Misra, Morikawa, Radford, Knight,
  Brundage, Murati, Mayer, Welinder, McGrew, Amodei, McCandlish, Sutskever, and
  Zaremba}]{Chen2021EvaluatingLL}
Mark Chen, Jerry Tworek, Heewoo Jun, Qiming Yuan, Henrique Pond{\'e}, Jared
  Kaplan, Harrison Edwards, Yura Burda, Nicholas Joseph, Greg Brockman, Alex
  Ray, Raul Puri, Gretchen Krueger, Michael Petrov, Heidy Khlaaf, Girish
  Sastry, Pamela Mishkin, Brooke Chan, Scott Gray, and 34 others. 2021.
\newblock \href {https://api.semanticscholar.org/CorpusID:235755472}
  {Evaluating large language models trained on code}.
\newblock \emph{ArXiv}, abs/2107.03374.

\bibitem[{Collberg and Sahoo(2005)}]{collberg2005software}
Christian Collberg and Tapas~Ranjan Sahoo. 2005.
\newblock Software watermarking in the frequency domain: implementation,
  analysis, and attacks.
\newblock \emph{Journal of Computer Security}, 13(5):721--755.

\bibitem[{Collberg and Thomborson(1999)}]{collberg1999software}
Christian Collberg and Clark Thomborson. 1999.
\newblock Software watermarking: Models and dynamic embeddings.
\newblock In \emph{Proceedings of the 26th ACM SIGPLAN-SIGACT symposium on
  Principles of programming languages}, pages 311--324.

\bibitem[{Dathathri et~al.(2024)Dathathri, See, Ghaisas, Huang, McAdam, Welbl,
  Bachani, Kaskasoli, Stanforth, Matejovicova et~al.}]{dathathri2024scalable}
Sumanth Dathathri, Abigail See, Sumedh Ghaisas, Po-Sen Huang, Rob McAdam,
  Johannes Welbl, Vandana Bachani, Alex Kaskasoli, Robert Stanforth, Tatiana
  Matejovicova, et~al. 2024.
\newblock Scalable watermarking for identifying large language model outputs.
\newblock \emph{Nature}, 634(8035):818--823.

\bibitem[{Dey et~al.(2019)Dey, Bhattacharya, and Chaki}]{dey2019software}
Ayan Dey, Sukriti Bhattacharya, and Nabendu Chaki. 2019.
\newblock Software watermarking: Progress and challenges.
\newblock \emph{INAE Letters}, 4(1):65--75.

\bibitem[{Feng et~al.(2025)Feng, Zhang, Zhang, Zhang, and Pan}]{feng2025bimark}
Xiaoyan Feng, He~Zhang, Yanjun Zhang, Leo~Yu Zhang, and Shirui Pan. 2025.
\newblock Bimark: Unbiased multilayer watermarking for large language models.
\newblock \emph{arXiv preprint arXiv:2506.21602}.

\bibitem[{Guan et~al.(2024)Guan, Wan, Bi, Wang, Zhang, Zhou, and
  Sun}]{guan2024codeip}
Batu Guan, Yao Wan, Zhangqian Bi, Zheng Wang, Hongyu Zhang, Pan Zhou, and
  Lichao Sun. 2024.
\newblock Codeip: A grammar-guided multi-bit watermark for large language
  models of code.
\newblock In \emph{Findings of the Association for Computational Linguistics:
  EMNLP 2024}, pages 9243--9258.

\bibitem[{Hanley and McNeil(1982)}]{hanley1982meaning}
James~A Hanley and Barbara~J McNeil. 1982.
\newblock The meaning and use of the area under a receiver operating
  characteristic (roc) curve.
\newblock \emph{Radiology}, 143(1):29--36.

\bibitem[{Hui et~al.(2024)Hui, Yang, Cui, Yang, Liu, Zhang, Liu, Zhang, Yu, Lu
  et~al.}]{hui2024qwen2}
Binyuan Hui, Jian Yang, Zeyu Cui, Jiaxi Yang, Dayiheng Liu, Lei Zhang, Tianyu
  Liu, Jiajun Zhang, Bowen Yu, Keming Lu, et~al. 2024.
\newblock Qwen2.5-coder technical report.
\newblock \emph{arXiv preprint arXiv:2409.12186}.

\bibitem[{Huo et~al.(2024)Huo, Somayajula, Liang, Zhang, Koushanfar, and
  Xie}]{huo2024token}
Mingjia Huo, Sai~Ashish Somayajula, Youwei Liang, Ruisi Zhang, Farinaz
  Koushanfar, and Pengtao Xie. 2024.
\newblock Token-specific watermarking with enhanced detectability and semantic
  coherence for large language models.
\newblock \emph{arXiv preprint arXiv:2402.18059}.

\bibitem[{Kim et~al.(2026)Kim, Park, and Han}]{kim2026marking}
Jungin Kim, Shinwoo Park, and Yo-Sub Han. 2026.
\newblock Marking code without breaking it: Code watermarking for detecting
  llm-generated code.
\newblock In \emph{Findings of the Association for Computational Linguistics:
  EACL 2026}, pages 3990--4002.

\bibitem[{Kirchenbauer et~al.(2023{\natexlab{a}})Kirchenbauer, Geiping, Wen,
  Katz, Miers, and Goldstein}]{kirchenbauer2023watermark}
John Kirchenbauer, Jonas Geiping, Yuxin Wen, Jonathan Katz, Ian Miers, and Tom
  Goldstein. 2023{\natexlab{a}}.
\newblock A watermark for large language models.
\newblock In \emph{International Conference on Machine Learning}, pages
  17061--17084. PMLR.

\bibitem[{Kirchenbauer et~al.(2023{\natexlab{b}})Kirchenbauer, Geiping, Wen,
  Shu, Saifullah, Kong, Fernando, Saha, Goldblum, and
  Goldstein}]{kirchenbauer2023reliability}
John Kirchenbauer, Jonas Geiping, Yuxin Wen, Manli Shu, Khalid Saifullah, Kezhi
  Kong, Kasun Fernando, Aniruddha Saha, Micah Goldblum, and Tom Goldstein.
  2023{\natexlab{b}}.
\newblock On the reliability of watermarks for large language models.
\newblock \emph{arXiv preprint arXiv:2306.04634}.

\bibitem[{Lee et~al.(2024)Lee, Hong, Ahn, Hong, Lee, Yun, Shin, and
  Kim}]{lee2024wrote}
Taehyun Lee, Seokhee Hong, Jaewoo Ahn, Ilgee Hong, Hwaran Lee, Sangdoo Yun,
  Jamin Shin, and Gunhee Kim. 2024.
\newblock Who wrote this code? watermarking for code generation.
\newblock In \emph{Proceedings of the 62nd Annual Meeting of the Association
  for Computational Linguistics (Volume 1: Long Papers)}, pages 4890--4911.

\bibitem[{Li et~al.(2023{\natexlab{a}})Li, Chen, Tang, Huang, Zhang, Zhang, and
  Yu}]{li2023functionmarker}
Shuai Li, Kejiang Chen, Kunsheng Tang, Wen Huang, Jie Zhang, Weiming Zhang, and
  Nenghai Yu. 2023{\natexlab{a}}.
\newblock Functionmarker: Watermarking language datasets via knowledge
  injection.
\newblock \emph{arXiv preprint arXiv:2311.09535}.

\bibitem[{Li et~al.(2023{\natexlab{b}})Li, Wang, Wang, and
  Gao}]{li2023protecting}
Zongjie Li, Chaozheng Wang, Shuai Wang, and Cuiyun Gao. 2023{\natexlab{b}}.
\newblock Protecting intellectual property of large language model-based code
  generation apis via watermarks.
\newblock In \emph{Proceedings of the 2023 ACM SIGSAC Conference on Computer
  and Communications Security}, pages 2336--2350.

\bibitem[{Liu et~al.(2024{\natexlab{a}})Liu, Pan, Hu, Li, Wen, King, and
  Yu}]{liu2024unforgeable}
Aiwei Liu, Leyi Pan, Xuming Hu, Shuang Li, Lijie Wen, Irwin King, and Philip
  Yu. 2024{\natexlab{a}}.
\newblock An unforgeable publicly verifiable watermark for large language
  models.
\newblock In \emph{International Conference on Learning Representations},
  volume 2024, pages 8291--8307.

\bibitem[{Liu et~al.(2024{\natexlab{b}})Liu, Pan, Hu, Meng, and
  Wen}]{liu2024semantic}
Aiwei Liu, Leyi Pan, Xuming Hu, Shiao Meng, and Lijie Wen. 2024{\natexlab{b}}.
\newblock A semantic invariant robust watermark for large language models.
\newblock In \emph{International Conference on Learning Representations},
  volume 2024, pages 6499--6519.

\bibitem[{Liu et~al.(2023)Liu, Xia, Wang, and Zhang}]{evalplus}
Jiawei Liu, Chunqiu~Steven Xia, Yuyao Wang, and Lingming Zhang. 2023.
\newblock \href {https://openreview.net/forum?id=1qvx610Cu7} {Is your code
  generated by chat{GPT} really correct? rigorous evaluation of large language
  models for code generation}.
\newblock In \emph{Thirty-seventh Conference on Neural Information Processing
  Systems}.

\bibitem[{Liu et~al.(2024{\natexlab{c}})Liu, Xie, Wang, Wei, Ding, and
  Zhang}]{evalperf}
Jiawei Liu, Songrun Xie, Junhao Wang, Yuxiang Wei, Yifeng Ding, and Lingming
  Zhang. 2024{\natexlab{c}}.
\newblock \href {https://openreview.net/forum?id=IBCBMeAhmC} {Evaluating
  language models for efficient code generation}.
\newblock In \emph{First Conference on Language Modeling}.

\bibitem[{Lozhkov et~al.(2024)Lozhkov, Li, Allal, Cassano, Lamy-Poirier, Tazi,
  Tang, Pykhtar, Liu, Wei et~al.}]{lozhkov2024starcoder}
Anton Lozhkov, Raymond Li, Loubna~Ben Allal, Federico Cassano, Joel
  Lamy-Poirier, Nouamane Tazi, Ao~Tang, Dmytro Pykhtar, Jiawei Liu, Yuxiang
  Wei, et~al. 2024.
\newblock Starcoder 2 and the stack v2: The next generation.
\newblock \emph{arXiv preprint arXiv:2402.19173}.

\bibitem[{Lu et~al.(2024)Lu, Liu, Yu, Li, and King}]{lu2024entropy}
Yijian Lu, Aiwei Liu, Dianzhi Yu, Jingjing Li, and Irwin King. 2024.
\newblock An entropy-based text watermarking detection method.
\newblock \emph{arXiv preprint arXiv:2403.13485}.

\bibitem[{Mao et~al.(2024)Mao, Wei, Chen, Fang, and Chau}]{mao2024watermarking}
Minjia Mao, Dongjun Wei, Zeyu Chen, Xiao Fang, and Michael Chau. 2024.
\newblock Watermarking low-entropy generation for large language models: An
  unbiased and low-risk method.
\newblock \emph{arXiv preprint arXiv:2405.14604}.

\bibitem[{Muennighoff et~al.(2023)Muennighoff, Liu, Liu, Zebaze, Zheng, Hui,
  Zhuo, Singh, Tang, von Werra, and Longpre}]{Muennighoff2023OctoPackIT}
Niklas Muennighoff, Qian Liu, Qi~Liu, Armel~Randy Zebaze, Qinkai Zheng, Binyuan
  Hui, Terry~Yue Zhuo, Swayam Singh, Xiangru Tang, Leandro von Werra, and
  S.~Longpre. 2023.
\newblock \href {https://api.semanticscholar.org/CorpusID:260886874} {Octopack:
  Instruction tuning code large language models}.
\newblock \emph{ArXiv}, abs/2308.07124.

\bibitem[{Pan et~al.(2024)Pan, Liu, He, Gao, Zhao, Lu, Zhou, Liu, Hu, Wen,
  King, and Yu}]{pan-etal-2024-markllm}
Leyi Pan, Aiwei Liu, Zhiwei He, Zitian Gao, Xuandong Zhao, Yijian Lu, Binglin
  Zhou, Shuliang Liu, Xuming Hu, Lijie Wen, Irwin King, and Philip~S. Yu. 2024.
\newblock \href {https://doi.org/10.18653/v1/2024.emnlp-demo.7} {{M}ark{LLM}:
  An open-source toolkit for {LLM} watermarking}.
\newblock In \emph{Proceedings of the 2024 Conference on Empirical Methods in
  Natural Language Processing: System Demonstrations}, pages 61--71, Miami,
  Florida, USA. Association for Computational Linguistics.

\bibitem[{Qu et~al.(2025)Qu, Zheng, Tao, Yin, Jiang, Tian, Zou, Jia, and
  Zhang}]{qu2025provably}
Wenjie Qu, Wengrui Zheng, Tianyang Tao, Dong Yin, Yanze Jiang, Zhihua Tian, Wei
  Zou, Jinyuan Jia, and Jiaheng Zhang. 2025.
\newblock Provably robust multi-bit watermarking for $\{$AI-generated$\}$ text.
\newblock In \emph{34th USENIX Security Symposium (USENIX Security 25)}, pages
  201--220.

\bibitem[{Ren et~al.(2024)Ren, Xu, Liu, Cui, Wang, Yin, and
  Tang}]{ren2024robust}
Jie Ren, Han Xu, Yiding Liu, Yingqian Cui, Shuaiqiang Wang, Dawei Yin, and
  Jiliang Tang. 2024.
\newblock A robust semantics-based watermark for large language model against
  paraphrasing.
\newblock In \emph{Findings of the Association for Computational Linguistics:
  NAACL 2024}, pages 613--625.

\bibitem[{Sato et~al.(2023)Sato, Takezawa, Bao, Niwa, and
  Yamada}]{sato2023embarrassingly}
Ryoma Sato, Yuki Takezawa, Han Bao, Kenta Niwa, and Makoto Yamada. 2023.
\newblock Embarrassingly simple text watermarks.
\newblock \emph{arXiv preprint arXiv:2310.08920}.

\bibitem[{Suresh et~al.(2024)Suresh, Ugare, Singh, and
  Misailovic}]{suresh2024watermarking}
Tarun Suresh, Shubham Ugare, Gagandeep Singh, and Sasa Misailovic. 2024.
\newblock Is the watermarking of llm-generated code robust?
\newblock \emph{arXiv preprint arXiv:2403.17983}.

\bibitem[{Takezawa et~al.(2023)Takezawa, Sato, Bao, Niwa, and
  Yamada}]{takezawa2023necessary}
Yuki Takezawa, Ryoma Sato, Han Bao, Kenta Niwa, and Makoto Yamada. 2023.
\newblock Necessary and sufficient watermark for large language models.
\newblock \emph{arXiv preprint arXiv:2310.00833}.

\bibitem[{Wang et~al.(2023)Wang, Yang, Chen, Zhou, Lin, Meng, Zhou, and
  Sun}]{wang2023towards}
Lean Wang, Wenkai Yang, Deli Chen, Hao Zhou, Yankai Lin, Fandong Meng, Jie
  Zhou, and Xu~Sun. 2023.
\newblock Towards codable watermarking for injecting multi-bits information to
  llms.
\newblock \emph{arXiv preprint arXiv:2307.15992}.

\bibitem[{Wang et~al.(2025)Wang, Gu, Wu, and Yang}]{wang2025morphmark}
Zongqi Wang, Tianle Gu, Baoyuan Wu, and Yujiu Yang. 2025.
\newblock Morphmark: Flexible adaptive watermarking for large language models.
\newblock In \emph{Proceedings of the 63rd Annual Meeting of the Association
  for Computational Linguistics (Volume 1: Long Papers)}, pages 4842--4860.

\bibitem[{Yang et~al.(2025)Yang, Li, Yang, Zhang, Hui, Zheng, Yu, Gao, Huang,
  Lv, Zheng, Liu, Zhou, Huang, Hu, Ge, Wei, Lin, Tang, Yang, Tu, Zhang, Yang,
  Yang, Zhou, Zhou, Lin, Dang, Bao, Yang, Yu, Deng, Li, Xue, Li, Zhang, Wang,
  Zhu, Men, Gao, Liu, Luo, Li, Tang, Yin, Ren, Wang, Zhang, Ren, Fan, Su,
  Zhang, Zhang, Wan, Liu, Wang, Cui, Zhang, Zhou, and Qiu}]{Yang2025Qwen3TR}
An~Yang, Anfeng Li, Baosong Yang, Beichen Zhang, Binyuan Hui, Bo~Zheng, Bowen
  Yu, Chang Gao, Chengen Huang, Chenxu Lv, Chujie Zheng, Dayiheng Liu, Fan
  Zhou, Fei Huang, Feng Hu, Hao Ge, Haoran Wei, Huan Lin, Jialong Tang, and 41
  others. 2025.
\newblock \href {https://api.semanticscholar.org/CorpusID:278602855} {Qwen3
  technical report}.
\newblock \emph{ArXiv}, abs/2505.09388.

\bibitem[{Yao et~al.(2024)Yao, Lou, Qin, and Ren}]{yao2024promptcare}
Hongwei Yao, Jian Lou, Zhan Qin, and Kui Ren. 2024.
\newblock Promptcare: Prompt copyright protection by watermark injection and
  verification.
\newblock In \emph{2024 IEEE Symposium on Security and Privacy (SP)}, pages
  845--861. IEEE.

\bibitem[{Zhang et~al.(2024)Zhang, Hussain, Neekhara, and
  Koushanfar}]{zhang2024remark}
Ruisi Zhang, Shehzeen~Samarah Hussain, Paarth Neekhara, and Farinaz Koushanfar.
  2024.
\newblock $\{$REMARK-LLM$\}$: A robust and efficient watermarking framework for
  generative large language models.
\newblock In \emph{33rd USENIX Security Symposium (USENIX Security 24)}, pages
  1813--1830.

\bibitem[{Zhao et~al.(2023)Zhao, Ananth, Li, and Wang}]{zhao2023provable}
Xuandong Zhao, Prabhanjan Ananth, Lei Li, and Yu-Xiang Wang. 2023.
\newblock Provable robust watermarking for ai-generated text.
\newblock \emph{arXiv preprint arXiv:2306.17439}.

\end{thebibliography}

\newpage
\appendix

\section{Example Grammar Rules}
\label{Example Grammar Rules}
Tables~\ref{tab:cfg_rules_python}, \ref{tab:cfg_rules_java}, and \ref{tab:cfg_rules_go} provide examples of context-free grammar for Python, Java, and Go specified in extended Backus–Naur form (EBNF) rules, illustrating the syntactic constraints used by GDW to determine which candidate tokens are eligible to receive watermark bias during decoding.

\section{Structural Role Categorization}
\label{Structural Role Categorization}
To support role-aware watermark injection, we partition the vocabulary of each target programming language into two disjoint categories according to their syntactic functions: \emph{syntax-critical tokens}, which govern grammatical structure and program formation, and \emph{content-bearing tokens}, which carry the remaining semantic content. Tables~\ref{tab:Syntax_Critical_tokens_python}, \ref{tab:Syntax_Critical_tokens_java}, and \ref{tab:Syntax_Critical_tokens_go} provide the comprehensive taxonomies of syntax-critical tokens for Python, Java, and Go.

\begin{table}[t]
\centering
\small
\setlength{\tabcolsep}{3pt}
\renewcommand{\arraystretch}{1.12}
\caption{Example Grammar Rules (EBNF) in Python}
\label{tab:cfg_rules_python}

\resizebox{\columnwidth}{!}{%
\begin{tabular}{@{}ll@{}}
\toprule
\StructCell{\textbf{Structure}}
&
\HeaderCell{\textbf{EBNF Grammar Rules}}
\\
\midrule

\StructCell{Definitions}
&
\EBNFRule{funcdef:}
{"def" NAME "(" parameters? ")" ["->" test] ":"
(suite \textbar{} \_NL)}
\\

&
\EBNFRule{classdef:}
{"class" NAME ["(" [arguments] ")"] ":" suite}
\\

\midrule

\StructCell{Flow Control}
&
\EBNFRule{if\_stmt:}
{"if" test ":" suite ("elif" test ":" suite)*
["else" ":" suite]}
\\

&
\EBNFRule{while\_stmt:}
{"while" test ":" suite ["else" ":" suite]}
\\

&
\EBNFRule{for\_stmt:}
{"for" exprlist "in" testlist ":" suite
["else" ":" suite]}
\\

&
\EBNFRule{with\_stmt:}
{"with" with\_item ("," with\_item)* ":" suite}
\\

\midrule

\StructCell{Terminals}
&
\EBNFRule{NAME:}
{/[a-zA-Z\_]\textbackslash{}w*/}
\\

&
\EBNFRule{COMMENT:}
{/\#.*(\textbackslash{}n[\textbackslash{}t ]*)+/
\textbar{} LONG\_STRING}
\\

\bottomrule
\end{tabular}%
}
\end{table}

\begin{table}[t]
\centering
\small
\setlength{\tabcolsep}{3pt}
\renewcommand{\arraystretch}{1.12}
\caption{Example Grammar Rules (EBNF) in Java}
\label{tab:cfg_rules_java}

\resizebox{\columnwidth}{!}{%
\begin{tabular}{@{}ll@{}}
\toprule
\StructCell{\textbf{Structure}}
&
\HeaderCell{\textbf{EBNF Grammar Rules}}
\\
\midrule

\StructCell{Program Structure}
&
\EBNFRule{package\_declaration:}
{"package" name ";"}
\\

&
\EBNFRule{import\_declaration:}
{"import" name ("." "*")? ";"}
\\

\midrule

\StructCell{Definitions}
&
\EBNFRule{class\_declaration:}
{modifiers? "class" CNAME super? interfaces? class\_body}
\\

&
\EBNFRule{method\_declaration:}
{method\_header method\_body}
\\

&
\EBNFRule{method\_header:}
{modifiers? type\_parameters? type method\_declarator throws?}
\\

\midrule

\StructCell{Flow Control}
&
\EBNFRule{if\_then\_statement:}
{"if" "(" expression ")" statement}
\\

&
\EBNFRule{if\_then\_else\_statement:}
{"if" "(" expression ")" statement "else" statement}
\\

&
\EBNFRule{while\_statement:}
{"while" "(" expression ")" statement}
\\

&
\EBNFRule{for\_statement:}
{"for" "(" for\_each ")" statement
\textbar{} "for" "(" for\_traditional ")" statement}
\\

\midrule

\StructCell{Terminals}
&
\EBNFRule{boolean\_literal:}
{"true" \textbar{} "false"}
\\

&
\EBNFRule{null\_literal:}
{"null"}
\\

\bottomrule
\end{tabular}%
}
\end{table}

\begin{table}[t]
\centering
\small
\setlength{\tabcolsep}{3pt}
\renewcommand{\arraystretch}{1.12}
\caption{Example Grammar Rules (EBNF) in Go}
\label{tab:cfg_rules_go}

\resizebox{\columnwidth}{!}{%
\begin{tabular}{@{}ll@{}}
\toprule
\StructCell{\textbf{Structure}}
&
\HeaderCell{\textbf{EBNF Grammar Rules}}
\\
\midrule

\StructCell{Program Structure}
&
\EBNFRule{package\_clause:}
{package\_token NAME}
\\

&
\EBNFRule{import\_decl:}
{import\_token
(import\_spec \textbar{} lparen\_token
((import\_spec eos)* import\_spec eos?)? ")")}
\\

\midrule

\StructCell{Definitions}
&
\EBNFRule{function\_decl:}
{func\_token NAME type\_parameters? signature block?}
\\

&
\EBNFRule{method\_decl:}
{func\_token receiver NAME signature block?}
\\

&
\EBNFRule{type\_decl:}
{type\_token
(type\_spec \textbar{} lparen\_token
((type\_spec eos)* type\_spec eos?)? ")")}
\\

\midrule

\StructCell{Flow Control}
&
\EBNFRule{if\_stmt:}
{if\_token (simple\_stmt? eos)? expression block
(else\_token (if\_stmt \textbar{} block))?}
\\

&
\EBNFRule{switch\_stmt:}
{expr\_switch\_stmt \textbar{} type\_switch\_stmt}
\\

&
\EBNFRule{for\_stmt:}
{for\_token [for\_clause] block}
\\

\midrule

\StructCell{Terminals}
&
\EBNFRule{integer:}
{DECIMAL\_LIT \textbar{} BINARY\_LIT
\textbar{} OCTAL\_LIT \textbar{} HEX\_LIT}
\\

&
\EBNFRule{string\_:}
{RAW\_STRING\_LIT
\textbar{} INTERPRETED\_STRING\_LIT}
\\

\bottomrule
\end{tabular}%
}
\end{table}

\begin{table}[htbp]
\centering
\small
\caption{Syntax-Critical Tokens in Python}
\label{tab:Syntax_Critical_tokens_python}
\begin{tabular}{@{}p{1.0\linewidth}@{}}
\toprule
\textbf{Control Flow} (15) \\
\midrule
\texttt{if}, \texttt{elif}, \texttt{else}, \texttt{for}, \texttt{while}, \texttt{break}, \texttt{continue}, \texttt{return}, \texttt{yield}, \texttt{pass}, \texttt{raise}, \texttt{try}, \texttt{except}, \texttt{finally}, \texttt{with} \\
\midrule
\textbf{Definitions} (10) \\
\midrule
\texttt{class}, \texttt{def}, \texttt{lambda}, \texttt{del}, \texttt{global}, \texttt{nonlocal}, \texttt{import}, \texttt{from}, \texttt{as}, \texttt{assert} \\
\midrule
\textbf{Logic} (8) \\
\midrule
\texttt{and}, \texttt{or}, \texttt{not}, \texttt{in}, \texttt{is}, \texttt{True}, \texttt{False}, \texttt{None} \\
\midrule
\textbf{Async} (2) \\
\midrule
\texttt{async}, \texttt{await} \\
\midrule
\textbf{Built-in Types} (11) \\
\midrule
\texttt{bool}, \texttt{bytes}, \texttt{complex}, \texttt{dict}, \texttt{float}, \texttt{int}, \texttt{list}, \texttt{set}, \texttt{str}, \texttt{tuple}, \texttt{NoneType} \\
\midrule
\textbf{Operators} (27) \\
\midrule
\texttt{+}, \texttt{-}, \texttt{*}, \texttt{**}, \texttt{/}, \texttt{//}, \texttt{\%},  \texttt{=}, \texttt{==}, \texttt{!=}, \texttt{>}, \texttt{<}, \texttt{>=}, \texttt{<=}, \texttt{+=}, \texttt{-=}, \texttt{*=}, \texttt{/=}, \texttt{\%=}, \texttt{//=}, \texttt{**=}, \texttt{\&}, \texttt{|}, \texttt{\textless{}\textless}, \texttt{\textgreater{}\textgreater}, \texttt{\textasciicircum}, \texttt{\textasciitilde} \\
\midrule
\textbf{Delimiters} (13) \\
\midrule
\texttt{(}, \texttt{)}, \texttt{[}, \texttt{]}, \texttt{\{}, \texttt{\}}, \texttt{,}, \texttt{:}, \texttt{.}, \texttt{;}, \texttt{@}, \texttt{->}, \texttt{...} \\
\midrule
\textbf{Whitespace} (4) \\
\midrule
\texttt{space}, \texttt{\textbackslash{}t}, \texttt{\textbackslash{}n}, \texttt{\textbackslash{}r} \\
\bottomrule
\end{tabular}
\end{table}

\begin{table}[htbp]
\centering
\small
\caption{Syntax-Critical Tokens in Java}
\label{tab:Syntax_Critical_tokens_java}
\begin{tabular}{@{}p{1.0\linewidth}@{}}
\toprule
\textbf{Control Flow} (14) \\
\midrule
\texttt{if}, \texttt{else}, \texttt{switch}, \texttt{case}, \texttt{default}, \texttt{for}, \texttt{while}, \texttt{do}, \texttt{break}, \texttt{continue}, \texttt{return}, \texttt{try}, \texttt{catch}, \texttt{finally} \\
\midrule
\textbf{Definitions} (25) \\
\midrule
\texttt{class}, \texttt{interface}, \texttt{enum}, \texttt{extends}, \texttt{implements}, \texttt{package}, \texttt{import}, \texttt{public}, \texttt{private}, \texttt{protected}, \texttt{static}, \texttt{final}, \texttt{abstract}, \texttt{native}, \texttt{strictfp}, \texttt{transient}, \texttt{volatile}, \texttt{new}, \texttt{const}, \texttt{void}, \texttt{synchronized}, \texttt{assert}, \texttt{goto}, \texttt{throw}, \texttt{throws} \\
\midrule
\textbf{Logic} (9) \\
\midrule
\texttt{true}, \texttt{false}, \texttt{null}, \texttt{instanceof}, \texttt{this}, \texttt{super}, \texttt{\&\&}, \texttt{||}, \texttt{!} \\
\midrule
\textbf{Built-in Types} (10) \\
\midrule
\texttt{byte}, \texttt{short}, \texttt{int}, \texttt{long}, \texttt{float}, \texttt{double}, \texttt{boolean}, \texttt{char}, \texttt{String}, \texttt{Object} \\
\midrule
\textbf{Operators} (34) \\
\midrule
\texttt{+}, \texttt{-}, \texttt{*}, \texttt{/}, \texttt{\%}, \texttt{++}, \texttt{-{}-}, \texttt{=}, \texttt{==}, \texttt{!=}, \texttt{>}, \texttt{<}, \texttt{>=}, \texttt{<=}, \texttt{\&}, \texttt{|}, \texttt{\textasciicircum}, \texttt{\textasciitilde}, \texttt{\textless{}\textless}, \texttt{\textgreater{}\textgreater}, \texttt{\textgreater{}\textgreater{}\textgreater}, \texttt{+=}, \texttt{-=}, \texttt{*=}, \texttt{/=}, \texttt{\%=}, \texttt{\&=}, \texttt{|=}, \texttt{\textasciicircum=}, \texttt{\textless{}\textless=}, \texttt{\textgreater{}\textgreater=}, \texttt{\textgreater{}\textgreater{}\textgreater=}, \texttt{->}, \texttt{::} \\
\midrule
\textbf{Delimiters} (12) \\
\midrule
\texttt{(}, \texttt{)}, \texttt{[}, \texttt{]}, \texttt{\{}, \texttt{\}}, \texttt{,}, \texttt{:}, \texttt{.}, \texttt{;}, \texttt{@}, \texttt{...} \\
\midrule
\textbf{Whitespace} (3) \\
\midrule
\texttt{space}, \texttt{\textbackslash{}t}, \texttt{\textbackslash{}n} \\
\bottomrule
\end{tabular}
\end{table}

\begin{table}[htbp]
\centering
\small
\caption{Syntax-Critical Tokens in Go}
\label{tab:Syntax_Critical_tokens_go}
\begin{tabular}{@{}p{1.0\linewidth}@{}}
\toprule
\textbf{Control Flow} (12) \\
\midrule
\texttt{if}, \texttt{else}, \texttt{switch}, \texttt{case}, \texttt{default}, \texttt{for}, \texttt{range}, \texttt{break}, \texttt{continue}, \texttt{return}, \texttt{goto}, \texttt{fallthrough} \\
\midrule
\textbf{Definitions} (13) \\
\midrule
\texttt{func}, \texttt{package}, \texttt{import}, \texttt{type}, \texttt{const}, \texttt{var}, \texttt{struct}, \texttt{interface}, \texttt{map}, \texttt{defer}, \texttt{go}, \texttt{select}, \texttt{chan} \\
\midrule
\textbf{Logic} (7) \\
\midrule
\texttt{true}, \texttt{false}, \texttt{nil}, \texttt{iota}, \texttt{\&\&}, \texttt{||}, \texttt{!} \\
\midrule
\textbf{Built-in Types} (22) \\
\midrule
\texttt{bool}, \texttt{byte}, \texttt{rune}, \texttt{string}, \texttt{int}, \texttt{int8}, \texttt{int16}, \texttt{int32}, \texttt{int64}, \texttt{uint}, \texttt{uint8}, \texttt{uint16}, \texttt{uint32}, \texttt{uint64}, \texttt{uintptr}, \texttt{float32}, \texttt{float64}, \texttt{complex64}, \texttt{complex128}, \texttt{error}, \texttt{any}, \texttt{comparable} \\
\midrule
\textbf{Operators} (34) \\
\midrule
\texttt{+}, \texttt{-}, \texttt{*}, \texttt{/}, \texttt{\%}, \texttt{++}, \texttt{-{}-}, \texttt{=}, \texttt{==}, \texttt{!=}, \texttt{>}, \texttt{<}, \texttt{>=}, \texttt{<=}, \texttt{\&}, \texttt{|}, \texttt{\textasciicircum}, \texttt{\&\textasciicircum}, \texttt{\textasciitilde}, \texttt{\textless{}\textless}, \texttt{\textgreater{}\textgreater}, \texttt{<-}, \texttt{:=}, \texttt{+=}, \texttt{-=}, \texttt{*=}, \texttt{/=}, \texttt{\%=}, \texttt{\&=}, \texttt{|=}, \texttt{\textasciicircum=}, \texttt{\&\textasciicircum=}, \texttt{\textless{}\textless=}, \texttt{\textgreater{}\textgreater=}\\
\midrule
\textbf{Delimiters} (11) \\
\midrule
\texttt{(}, \texttt{)}, \texttt{[}, \texttt{]}, \texttt{\{}, \texttt{\}}, \texttt{,}, \texttt{;}, \texttt{:}, \texttt{.}, \texttt{...} \\
\midrule
\textbf{Whitespace} (4) \\
\midrule
\texttt{space}, \texttt{\textbackslash{}t}, \texttt{\textbackslash{}n}, \texttt{\textbackslash{}r} \\
\bottomrule
\end{tabular}
\end{table}




\section{Results on Qwen2.5-Coder-7B}
\label{Results on Qwen2.5-Coder-7B}

To further demonstrate the generalizability of GDW, we additionally conduct experiments on Qwen2.5-Coder-7B under beam search decoding. We evaluate GDW across three programming languages: Python on HumanEval dataset, Java and Go on HumanEvalPack dataset. 

Table~\ref{tab:autc_on_qwen7b} presents the AUTC results of different watermarking methods. GDW consistently achieves the best performance across all baselines. We also provide visualizations of the trade-off curves in Figure~\ref{fig:tradeoff_curves_on_qwen7b}, which show that GDW establishes a stronger quality-detectability trade-off frontier than existing methods. These experimental results demonstrate the strong generalizability of GDW across larger code generation models and multiple programming languages.

\begin{table}[t]
\centering
\normalsize
\setlength{\tabcolsep}{10pt}
\begin{tabular}{lccc}
\toprule
Methods & Python & Java & Go \\
\midrule
KGW   & 0.797 & 0.887 & 0.896 \\
SWEET & 0.765 & 0.800 & 0.844 \\
EWD   & 0.792 & 0.910 & 0.895 \\
STONE & 0.666 & 0.761 & -- \\
CodeIP & 0.738 & 0.801 & 0.839 \\
SynthID-Text & 0.687 & 0.705 & 0.701 \\
GDW   & \textbf{0.873} & \textbf{0.941} & \textbf{0.914} \\
\bottomrule
\end{tabular}
\caption{AUTC of different watermarking methods using Qwen2.5-Coder-7B with beam search decoding. The best result in each column is highlighted in bold.}
\label{tab:autc_on_qwen7b}
\end{table}

\begin{figure*}[t]
\centering
\setlength{\tabcolsep}{1pt}

\begin{subfigure}[t]{0.32\textwidth}
    \centering
    \includegraphics[width=\linewidth]{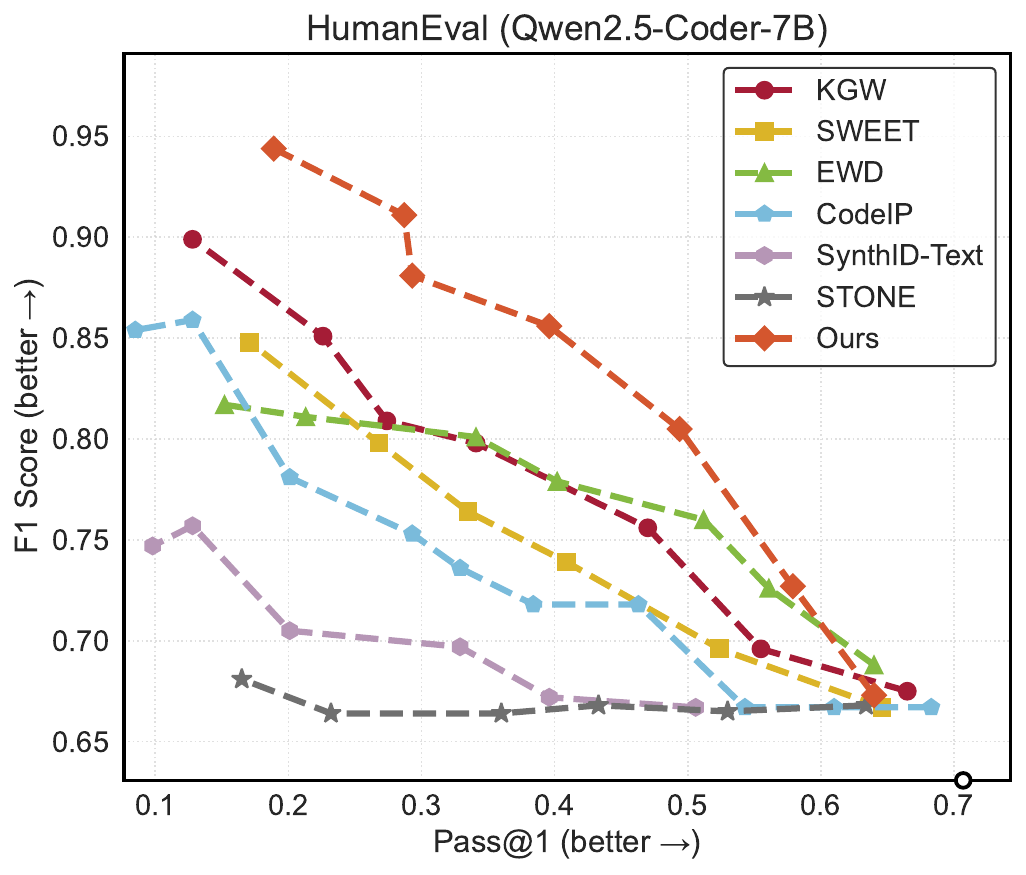}
\end{subfigure}
\hfill
\begin{subfigure}[t]{0.32\textwidth}
    \centering
    \includegraphics[width=\linewidth]{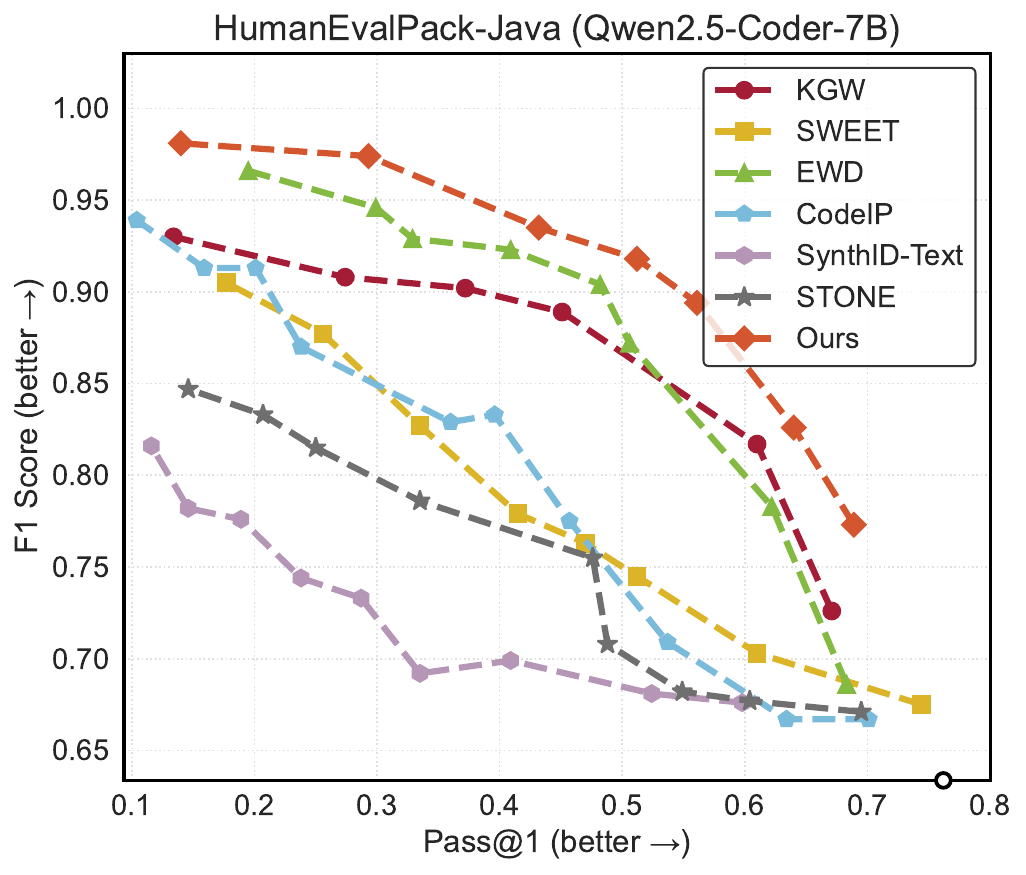}
\end{subfigure}
\hfill
\begin{subfigure}[t]{0.32\textwidth}
    \centering
    \includegraphics[width=\linewidth]{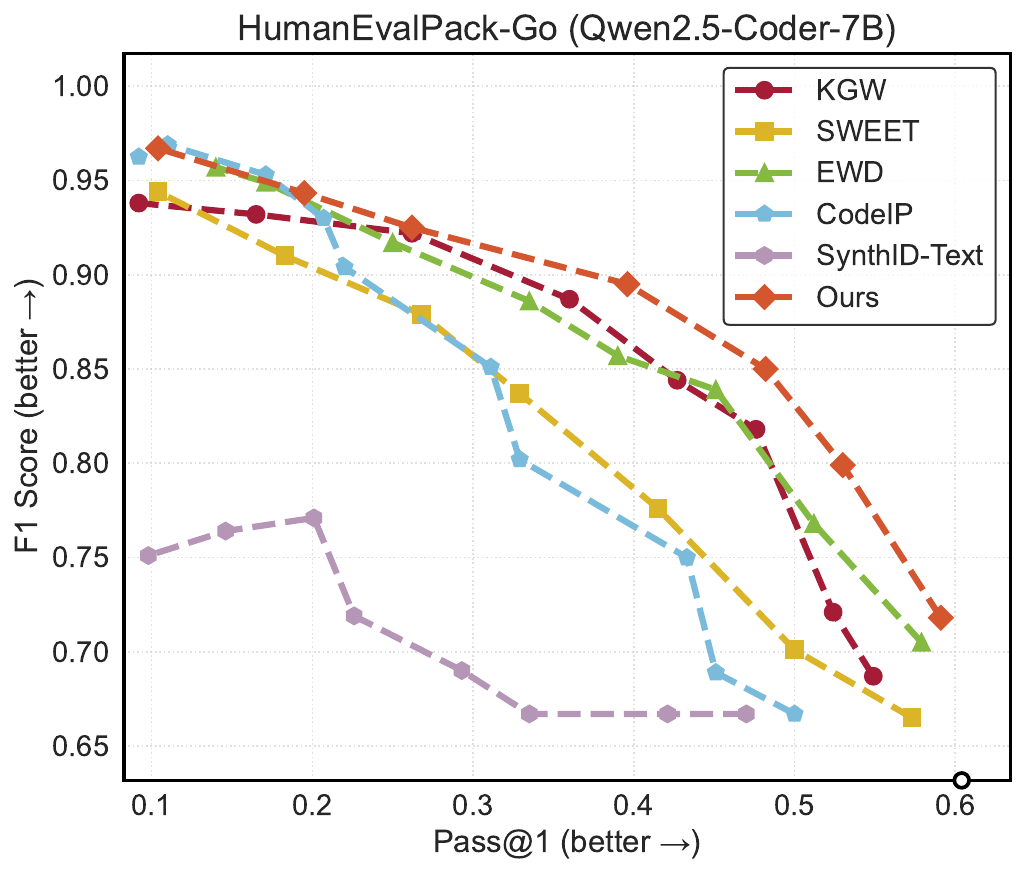}
\end{subfigure}

\caption{Pass@1-F1 trade-off curves using Qwen2.5-Coder-7B across three programming languages with beam search decoding. The hollow marker represents the unwatermarked pass@1 performance as a quality reference. GDW consistently achieves a more favorable quality-detectability trade-off than other methods.}

\label{fig:tradeoff_curves_on_qwen7b}
\end{figure*}

\begin{table*}[t]
\centering
\small
\setlength{\tabcolsep}{4.5pt}
\renewcommand{\arraystretch}{1.08}
\resizebox{\textwidth}{!}{
\begin{tabular}{llllc}
\toprule
\textbf{Experiment} & \textbf{Model} & \textbf{Benchmark} & \textbf{Setting} & \textbf{$[Q_{\text{start}}, Q_{\text{end}}]$} \\
\midrule

Python Evaluation
& Qwen2.5-Coder-3B & HumanEval & Sampling & [0.079, 0.390] \\
& Qwen2.5-Coder-3B & HumanEval & Beam Search & [0.134, 0.463] \\
& Qwen2.5-Coder-3B & MBPP+ & Sampling & [0.190, 0.569] \\
& Qwen2.5-Coder-3B & MBPP+ & Beam Search & [0.193, 0.561] \\
& Qwen3-4B & HumanEval & Sampling & [0.232, 0.537] \\
& Qwen3-4B & HumanEval & Beam Search & [0.256, 0.622] \\
& Qwen3-4B & MBPP+ & Sampling & [0.259, 0.566] \\
& Qwen3-4B & MBPP+ & Beam Search & [0.325, 0.566] \\
& StarCoder2-3B & HumanEval & Sampling & [0.128, 0.232] \\
& StarCoder2-3B & HumanEval & Beam Search & [0.110, 0.238] \\
& StarCoder2-3B & MBPP+ & Sampling & [0.217, 0.405] \\
& StarCoder2-3B & MBPP+ & Beam Search & [0.122, 0.429] \\
\midrule

Java and Go Evaluation
& Qwen2.5-Coder-3B & HumanEvalPack-Java & Sampling & [0.287, 0.561] \\
& Qwen2.5-Coder-3B & HumanEvalPack-Java & Beam Search & [0.140, 0.457] \\
& Qwen2.5-Coder-3B & HumanEvalPack-Go & Sampling & [0.140, 0.311] \\
& Qwen2.5-Coder-3B & HumanEvalPack-Go & Beam Search & [0.140, 0.323] \\
& StarCoder2-3B & HumanEvalPack-Java & Sampling & [0.183, 0.335] \\
& StarCoder2-3B & HumanEvalPack-Java & Beam Search & [0.171, 0.360] \\
& StarCoder2-3B & HumanEvalPack-Go & Sampling & [0.134, 0.213] \\
& StarCoder2-3B & HumanEvalPack-Go & Beam Search & [0.122, 0.244] \\
\midrule

Larger-Model Evaluation
& Qwen2.5-Coder-7B & HumanEval & Beam Search & [0.189, 0.506] \\
& Qwen2.5-Coder-7B & HumanEvalPack-Java & Beam Search & [0.195, 0.598] \\
& Qwen2.5-Coder-7B & HumanEvalPack-Go & Beam Search & [0.140, 0.470] \\
\midrule

Sensitivity Analysis of $\gamma$ 
& StarCoder2-3B & HumanEval & \begin{tabular}[t]{@{}l@{}}Beam Search \\ ($\gamma \in \{0.25, 0.50, 0.75\}$)\end{tabular} & [0.134, 0.226] \\
\midrule

Sensitivity Analysis of $\lambda$
& Qwen2.5-Coder-3B & HumanEvalPack-Java & \begin{tabular}[t]{@{}l@{}}Beam Search \\ ($\lambda \in \{2,3,4,5\}$)\end{tabular} & [0.104, 0.610] \\
\midrule

Robustness Analysis
& Qwen2.5-Coder-3B & HumanEval & \begin{tabular}[t]{@{}l@{}}Beam Search \\ (Seven  Attacks)\end{tabular} & [0.134, 0.396] \\
\midrule

Ablation Study
& StarCoder2-3B & HumanEval & \begin{tabular}[t]{@{}l@{}}Beam Search \\ (Three Ablated Variants)\end{tabular} & [0.091, 0.256] \\
\bottomrule

\end{tabular}
}
\caption{Common quality intervals $[Q_{\mathrm{start}}, Q_{\mathrm{end}}]$ used to compute AUTC under different experimental settings.}
\label{tab:common_ranges_autc}
\end{table*}

\begin{figure*}[t]
\centering
\setlength{\tabcolsep}{1pt}

\begin{subfigure}[t]{0.32\textwidth}
    \centering
    \includegraphics[width=\linewidth]{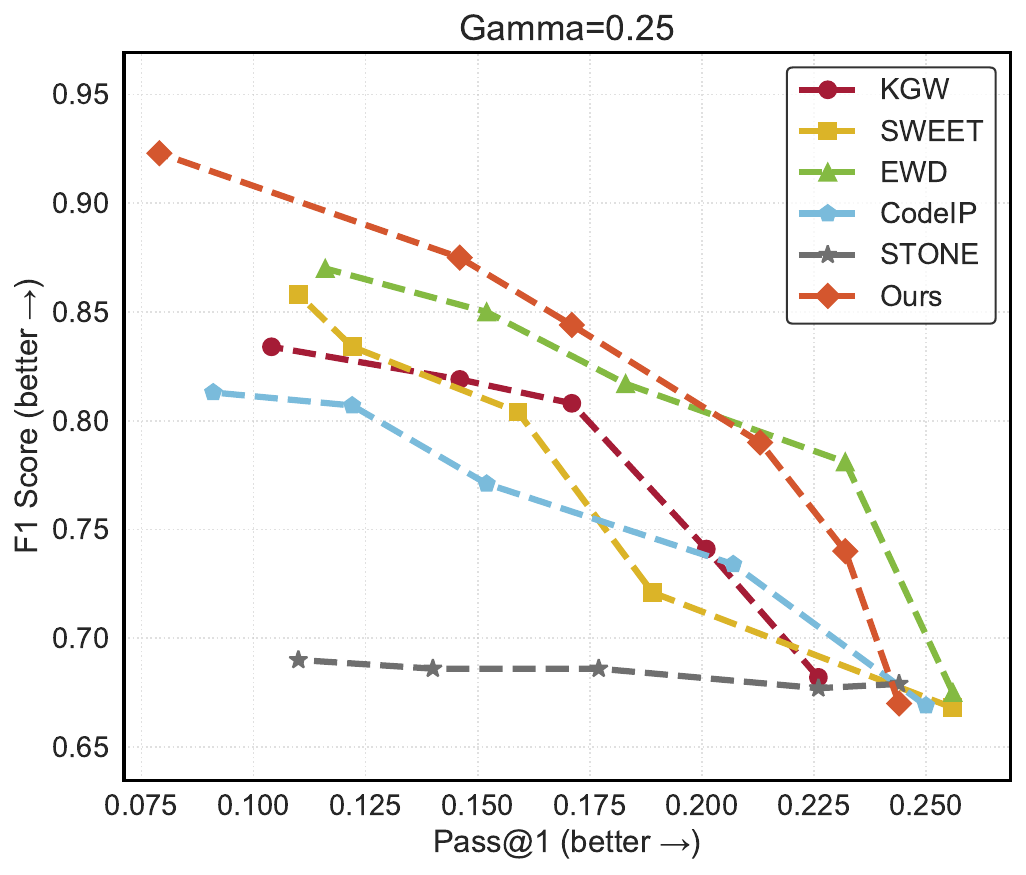}
\end{subfigure}
\hfill
\begin{subfigure}[t]{0.32\textwidth}
    \centering
    \includegraphics[width=\linewidth]{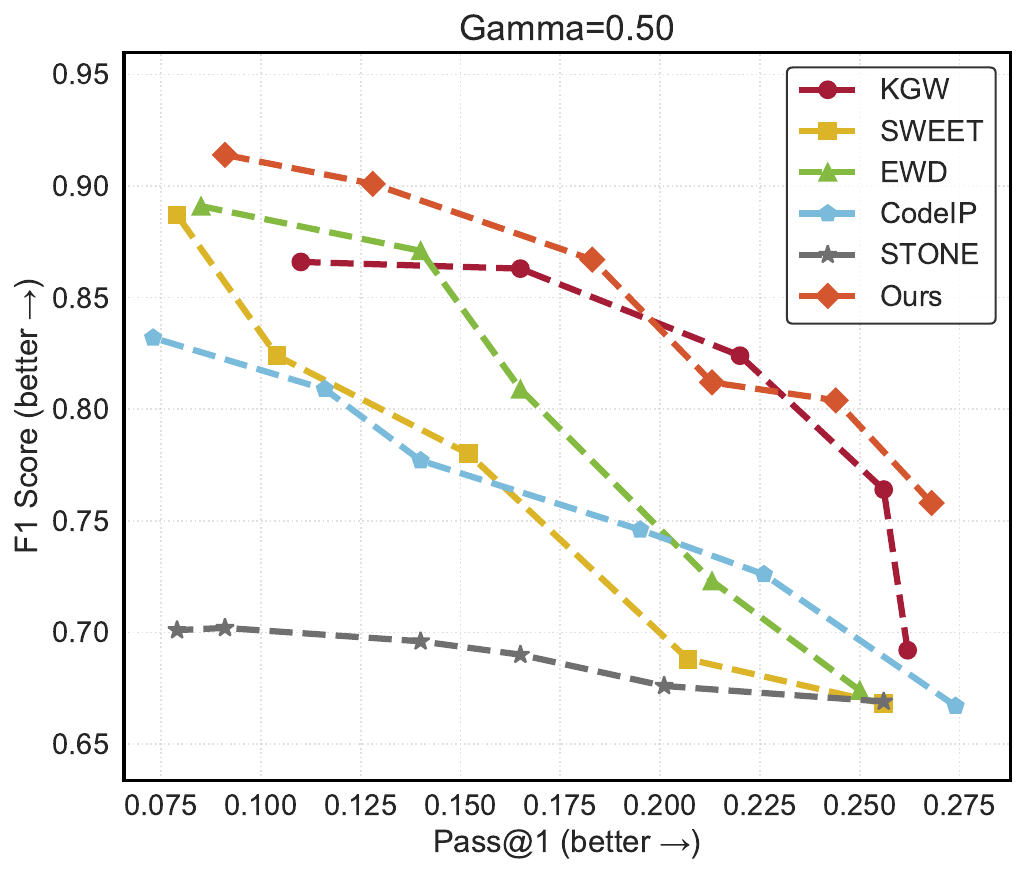}
\end{subfigure}
\hfill
\begin{subfigure}[t]{0.32\textwidth}
    \centering
    \includegraphics[width=\linewidth]{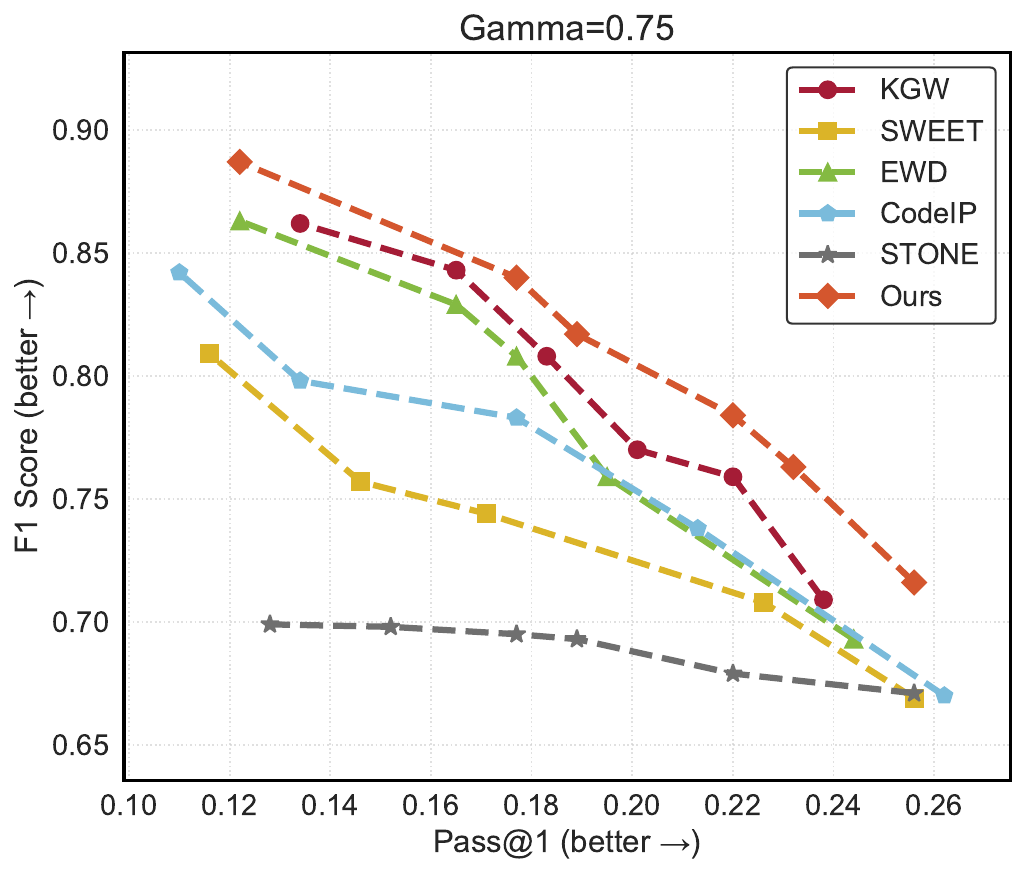}
\end{subfigure}
\caption{Pass@1-F1 trade-off curves under different values of the green-list ratio $\gamma$ on HumanEval using StarCoder2-3B with beam search decoding. GDW achieves stronger quality-detectability trade-offs under different $\gamma$ settings.}
\label{fig:gamma_comp}
\end{figure*}

\section{Common Quality Intervals Used for AUTC Computation}
\label{Common Quality Intervals Used for AUTC Computation}

To facilitate experimental reproducibility and fair comparison across watermarking methods, we report the common quality intervals used for AUTC computation in all experiments. These intervals correspond to $[Q_{\text{start}}, Q_{\text{end}}]$. Table~\ref{tab:common_ranges_autc} summarizes the common quality intervals used throughout all experiments in this paper.

\begin{table}[t]
\centering
\normalsize
\setlength{\tabcolsep}{5pt}
\begin{tabular}{lccc}
\toprule
Methods & $\gamma=0.25$ & $\gamma=0.50$ & $\gamma=0.75$ \\
\midrule
KGW   & 0.774 & 0.849 & 0.810 \\
SWEET & 0.754 & 0.734 & 0.738 \\
EWD   & 0.823 & 0.786 & 0.792 \\
STONE & 0.684 & 0.684 & 0.692 \\
CodeIP & 0.751 & 0.754 & 0.770 \\
SynthID-Text & -- & -- & -- \\
GDW   & \textbf{0.830} & \textbf{0.858} & \textbf{0.830} \\
\bottomrule
\end{tabular}
\caption{AUTC under different values of the green-list ratio $\gamma \in \{0.25, 0.50, 0.75\}$ on HumanEval using StarCoder2-3B with beam search decoding. All columns are computed over the same quality interval [0.134, 0.226].}
\label{tab:gamma_sensitivity_starcoder2_humaneval_beam}
\end{table}

\section{Discussion on AUTC Metric}
\label{Discussion on AUTC Metric}

To quantitatively evaluate the trade-off between code quality and watermark detectability, we introduce the Area Under the Trade-off Curve (AUTC), as formally defined in Eq.~\ref{eq:autc}. We provide additional details about it below.

\paragraph{Numerical Computation.}
For each method, we first sort the observed operating points by Pass@1 and retain all measured points, including locally non-monotonic ones. We then construct a piecewise-linear interpolation of detection F1 as a function of Pass@1. The interpolated curve is evaluated at 100 equally spaced quality levels over the common interval $[Q_{\mathrm{start}},Q_{\mathrm{end}}]$, and its area is approximated using the trapezoidal rule. Because the common interval is defined as the intersection of the observed Pass@1 ranges of all compared methods, all evaluation points lie within the observed range of every method. Finally, the resulting area is divided by $Q_{\mathrm{end}}-Q_{\mathrm{start}}$ to obtain AUTC.

\paragraph{Scope of AUTC Comparability.}
The normalization converts the unnormalized area into the mean interpolated detection F1 over the selected quality interval, thereby removing the area’s trivial linear dependence on interval width. However, AUTC still depends on the location and endpoints of the integration interval. Moreover, because the common quality interval \([Q_{\text{start}}, Q_{\text{end}}]\) is determined by the methods included in a comparison group, adding or removing a method may change the interval and consequently change the AUTC values of all methods in that group. Therefore, AUTC values are directly comparable only when they are computed over the identical quality interval. Our conclusions are based exclusively on comparisons among methods or variants evaluated within the same comparison group. The corresponding common quality intervals used for AUTC computation across different experimental settings are provided in Table~\ref{tab:common_ranges_autc}.

\paragraph{Comparison Between AUTC and STEM.}
We also consider the STEM metric introduced by STONE \citep{kim2026marking}, which aggregates Pass@k \citep{Chen2021EvaluatingLL}, perplexity \citep{brown1990statistical}, and AUROC \citep{hanley1982meaning} into a composite score at a single point. While STEM is useful for overall evaluation, such single-point metrics make it difficult to characterize the global trade-off behavior under different watermark strengths. 

In contrast, our goal is to evaluate the trade-off frontier between code quality and watermark detectability. By varying the watermark strength parameter, we obtain a set of Pass@1-F1 points that form a trade-off curve. AUTC summarizes the performance of the entire trade-off curve rather than a single point, making it more suitable for evaluating the quality-detectability trade-off. Notably, methods with similar STEM scores may still exhibit substantially different Pass@1-F1 trade-off frontiers, since STEM only evaluates a single operating point. This may also partially explain the relatively weaker AUTC performance of STONE in our experiments. Methodologically, the conservative code-oriented watermarking strategy adopted by STONE helps preserve code correctness by avoiding syntax-sensitive positions, but may also reduce the number of effective watermarkable tokens, thereby limiting watermark detectability under our evaluation setting. In contrast, GDW first limits watermark bias to grammatically admissible candidates and then applies stronger modulation to content-bearing tokens while maintaining conservative modulation for syntax-critical tokens, thereby achieving a better quality-detectability trade-off.

\section{Details of Experimental Settings}
\label{Details of Experimental Settings}
In this section, we report additional experimental details to ensure the reproducibility of our experiments. The experiments are conducted on NVIDIA RTX 5090 GPUs.

\textbf{Baseline Settings.}
Our proposed method and all baselines are implemented via the MarkLLM \citep{pan-etal-2024-markllm} toolkit. Note that CodeIP did not publicly release its key training code. Therefore, we reproduce its implementation based on the MarkLLM framework. The specific configurations for each method are detailed as follows:

\begin{itemize}
    \item \textbf{KGW} \citep{kirchenbauer2023watermark}:
    \raggedright
    $\gamma=0.5$, $\text{hash\_key}=15485863$, $\text{prefix\_length}=1$, $\text{f\_scheme}=\text{``time''}$, $\text{window\_scheme}=\text{``left''}$.
    \par

    \item \textbf{SWEET} \citep{lee2024wrote}:
    \raggedright
    $\gamma=0.5$, $\text{hash\_key}=15485863$, $\text{prefix\_length}=1$, $\text{entropy\_threshold}=0.7$.
    \par

    \item \textbf{EWD} \citep{lu2024entropy}:
    \raggedright
    $\gamma=0.5$, $\text{hash\_key}=15485863$, $\text{prefix\_length}=1$.
    \par

    \item \textbf{STONE} \citep{kim2026marking}:
    \raggedright
    $\gamma=0.5$, $\text{hash\_key}=15485863$, $\text{prefix\_length}=1$, $\text{skipping\_rule}=\text{``all\_pl''}$, $\text{watermark\_on\_pl}=\text{``False''}$, $\text{language}=\text{``python''}$.
    \par
    \textit{Note that STONE does not support Go, so the language field is restricted to \texttt{``python''} and \texttt{``java''}.}
    \par

    \item \textbf{CodeIP}:
    \raggedright
    $\gamma=0.5$, $\text{message\_code\_len}=10$, $\text{encode\_ratio}=0.5$, $\beta=0.5$, $\text{prompt\_length}=1024$, $\text{repeat\_penalty}=1.0$, $\text{ngram\_size}=0$, $\text{language}=\text{``python''}$.
    \par
    \textit{Note that CodeIP also supports Java and Go, so the language field can be set to \texttt{``python''}, \texttt{``java''} and \texttt{``go''}.}
    \par

    \item \textbf{SynthID-Text}:
    \raggedright
    $\text{ngram\_len}=5$, $\text{sampling\_table\_size}=65536$, $\text{sampling\_table\_seed}=0$, $\text{watermark\_mode}=\text{``distortionary''}$, $\text{num\_leaves}=5$, $\text{context\_history\_size}=128$, $\text{detector\_type}=\text{``bayesian''}$, $\text{threshold}=0.52$.
    \par
    \textit{Unlike red-green list based watermarking methods, we control SynthID-Text's watermark strength through the number of watermark keys. In our experiments, we use 30 watermark keys to maintain a balanced trade-off between code quality and detectability. The adopted keys are listed as follows: \seqsplit{[654, 400, 836, 123, 340, 443, 597, 160, 57, 29, 590, 639, 13, 715, 468, 990, 966, 226, 324, 585, 118, 504, 421, 521, 129, 669, 732, 225, 90, 960]}.}
    \par

    \item \textbf{GDW (Ours)}:
    \raggedright
    $\gamma=0.5$, $\text{hash\_key}=15485863$, $\text{prefix\_length}=1$, $\text{f\_scheme}=\text{``time''}$, $\text{window\_scheme}=\text{``left''}$, $\text{language}=\text{``python''}$.
    \par
    \textit{Note that for experiments on HumanEvalPack, the \texttt{language} field should be set to \texttt{``java''} and \texttt{``go''} respectively.}
    \par
\end{itemize}

\noindent\textbf{Decoding Strategy Settings.} 
For beam search, we set the beam width to $5$ with \verb|early_stopping| disabled ("never") to ensure stable and high-probability code generation. For sampling, we employ a temperature of $0.1$ and a $top\_p$ value of $0.95$ to balance diversity and functional correctness. For both strategies, the maximum number of newly generated tokens is fixed at $200$.

\section{Sensitivity to the Green-List Ratio $\gamma$}
\label{Sensitivity to the Green-List Ratio}

In the main experiments, we follow prior work and fix the green-list ratio $\gamma$ to 0.5. To assess the sensitivity of GDW to this choice, we additionally conduct experiments with $\gamma \in \{0.25, 0.50, 0.75\}$. To ensure direct comparability across $\gamma$ settings, we compute AUTC values over the same shared quality interval [0.134, 0.226]. As shown in Table~\ref{tab:gamma_sensitivity_starcoder2_humaneval_beam}, GDW achieves the highest AUTC under all three settings, which indicates that GDW's quality-detectability trade-off advantage does not depend on a single fixed green-list ratio. Figure~\ref{fig:gamma_comp} presents the corresponding Pass@1-F1 trade-off curves under different $\gamma$ settings.

\section{Supplementary Visualization of Trade-off Curves}
\label{Supplementary Visualization of Trade-off Curves}
\begin{figure*}[t]
\centering
\setlength{\tabcolsep}{1pt}

\begin{subfigure}[t]{0.24\textwidth}
    \centering
    \includegraphics[width=\linewidth]{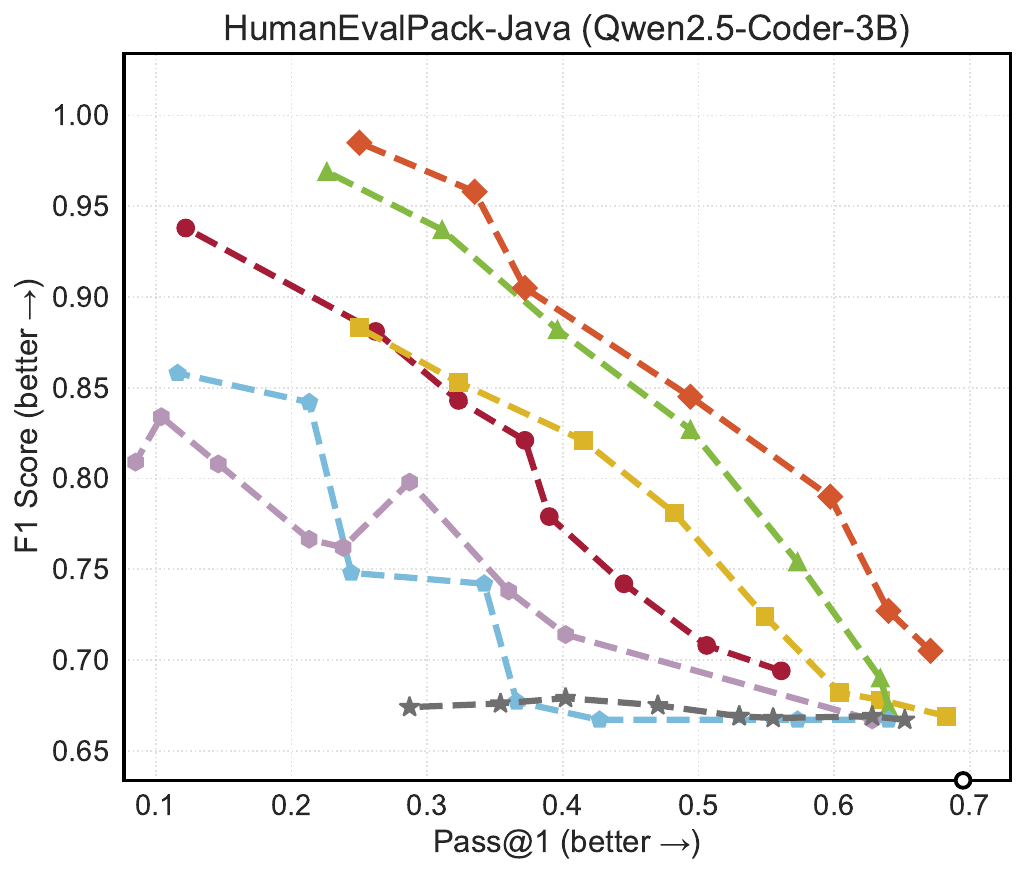}
\end{subfigure}
\hfill
\begin{subfigure}[t]{0.24\textwidth}
    \centering
    \includegraphics[width=\linewidth]{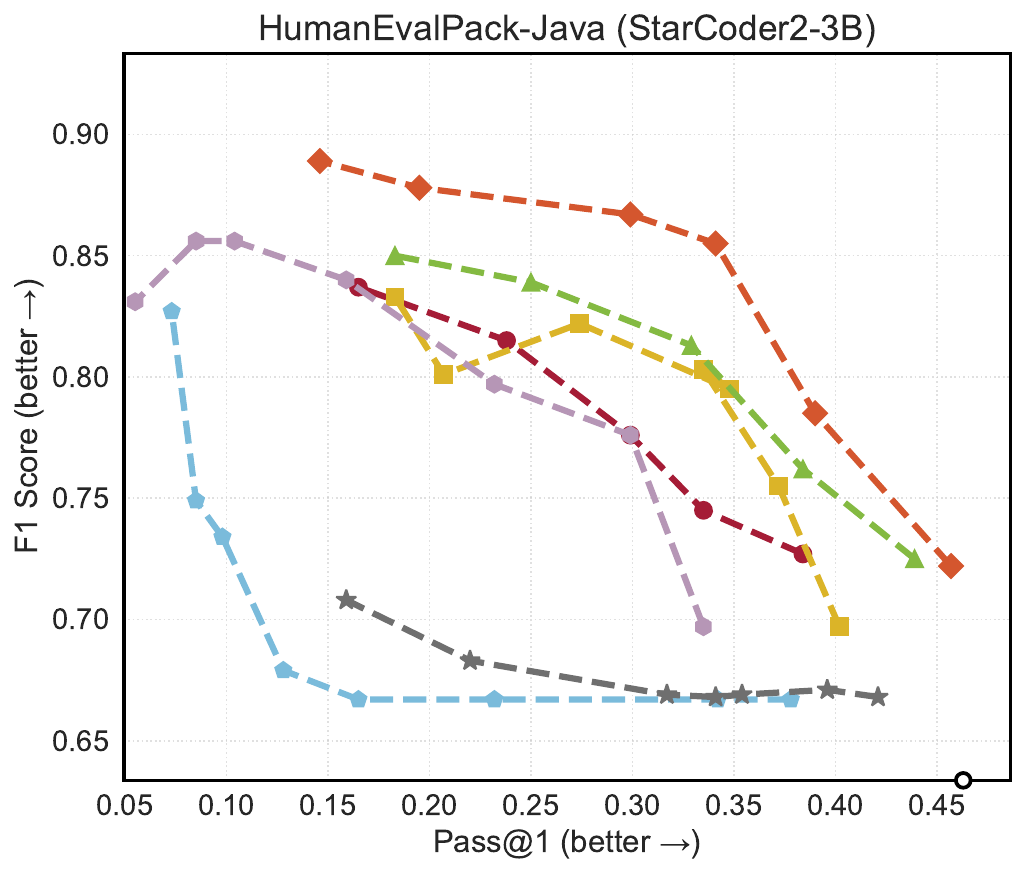}
\end{subfigure}
\hfill
\begin{subfigure}[t]{0.24\textwidth}
    \centering
    \includegraphics[width=\linewidth]{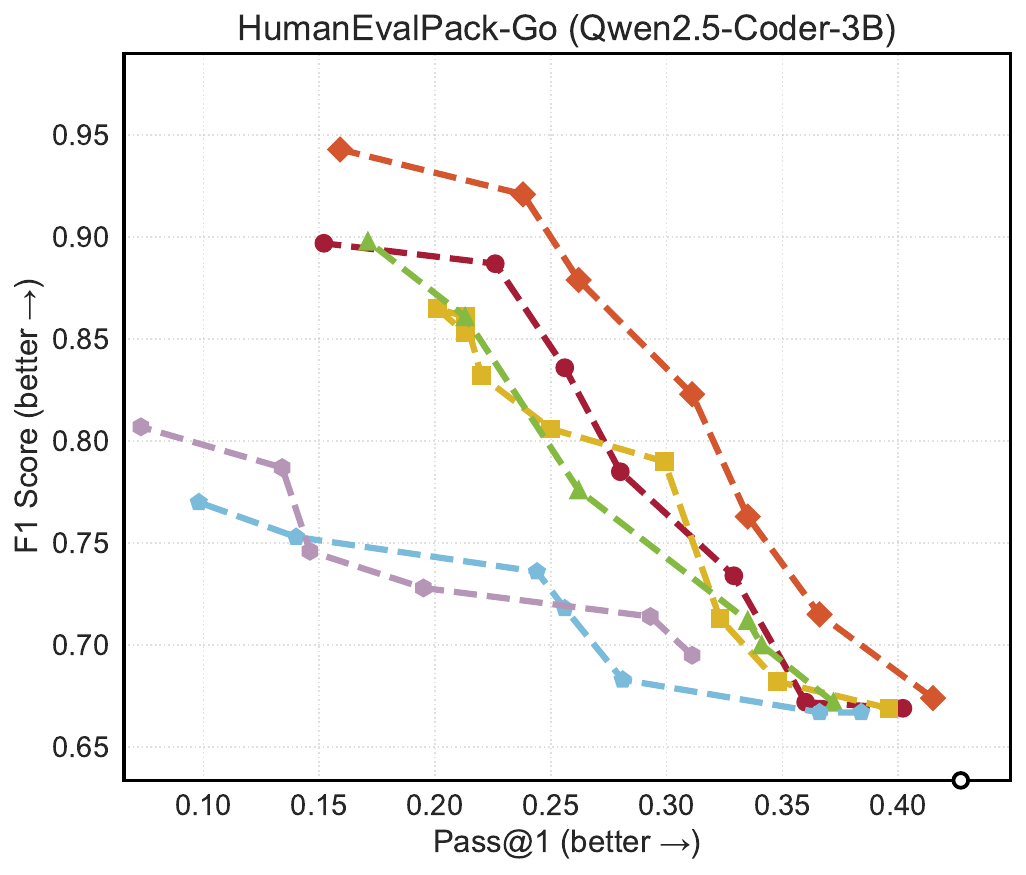}
\end{subfigure}
\hfill
\begin{subfigure}[t]{0.24\textwidth}
    \centering
    \includegraphics[width=\linewidth]{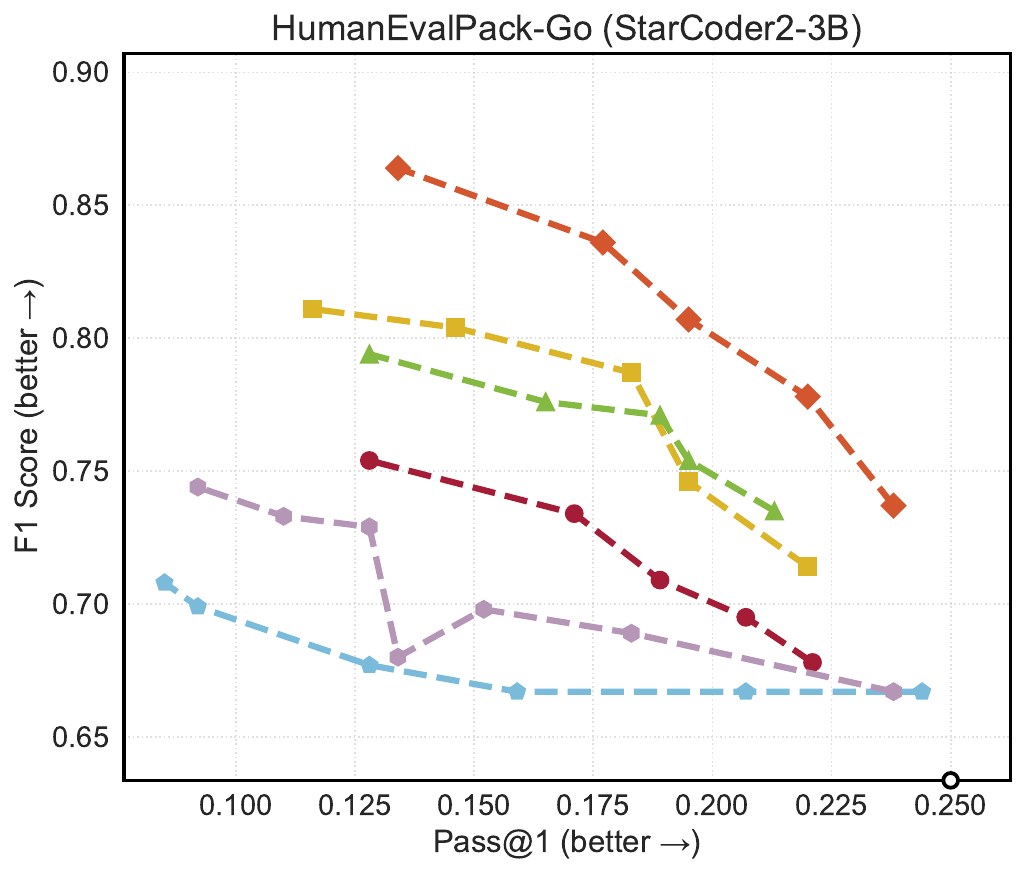}
\end{subfigure}

\begin{subfigure}[t]{0.24\textwidth}
    \centering
    \includegraphics[width=\linewidth]{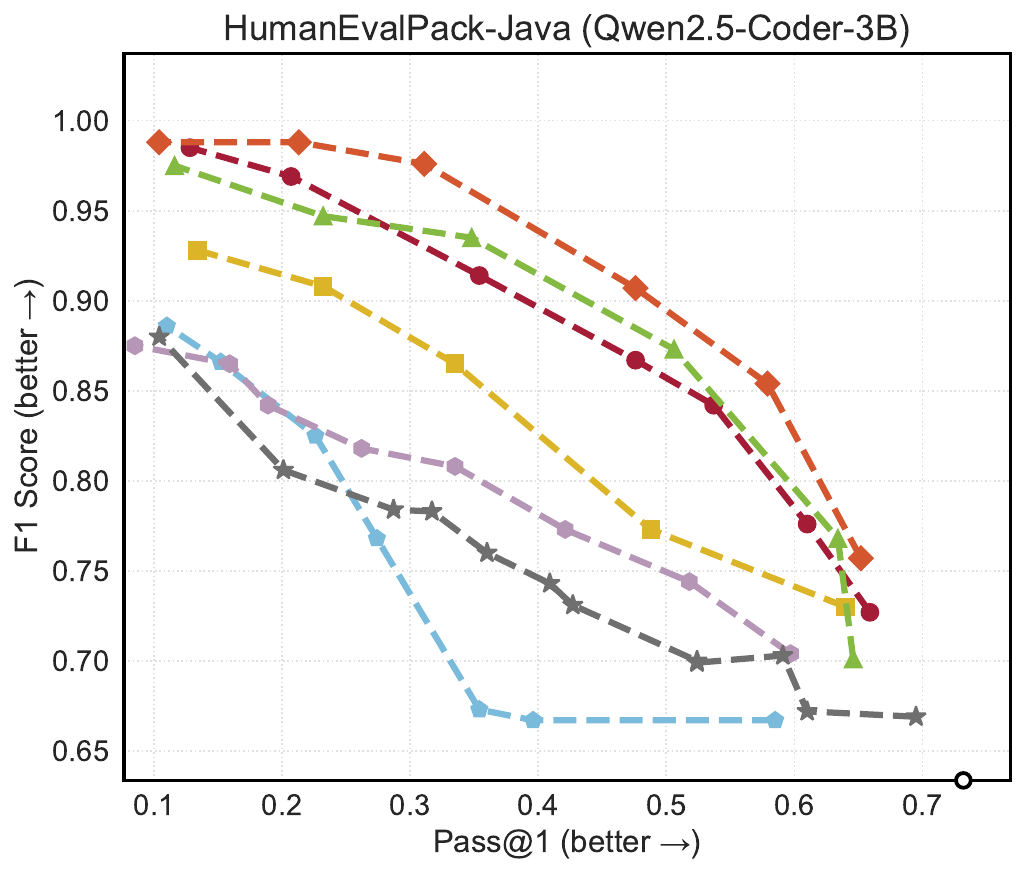}
\end{subfigure}
\hfill
\begin{subfigure}[t]{0.24\textwidth}
    \centering
    \includegraphics[width=\linewidth]{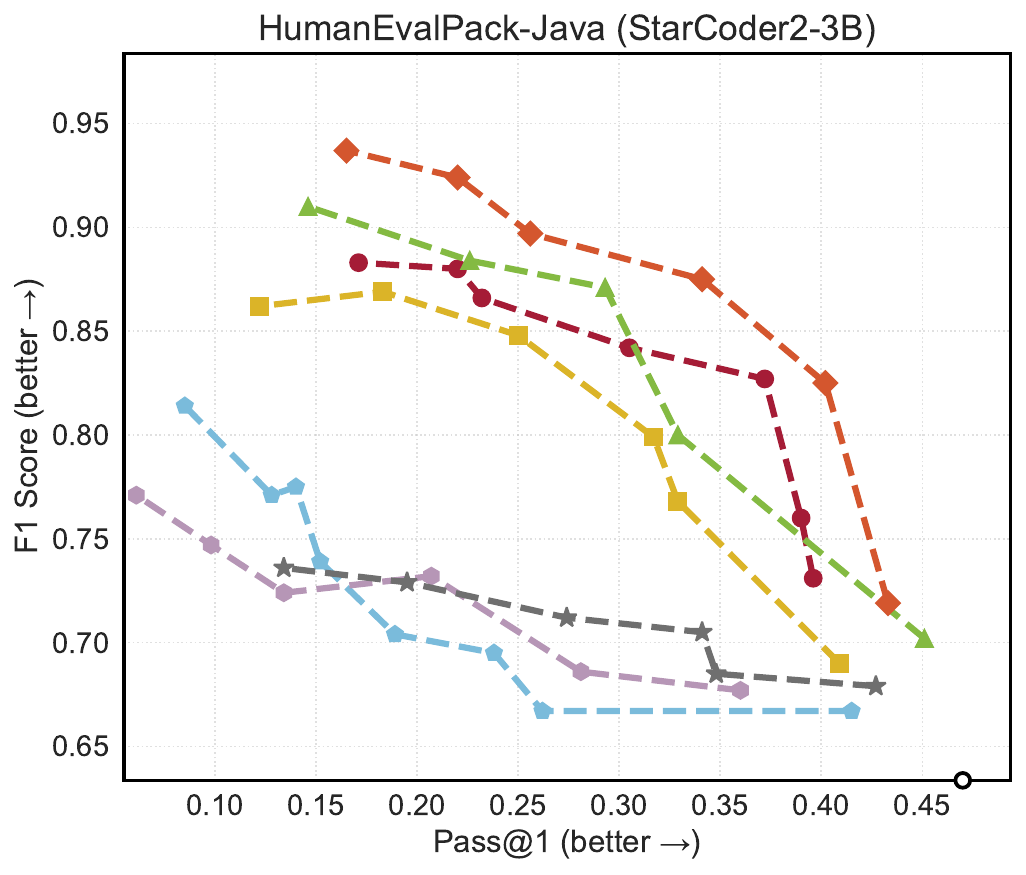}
\end{subfigure}
\hfill
\begin{subfigure}[t]{0.24\textwidth}
    \centering
    \includegraphics[width=\linewidth]{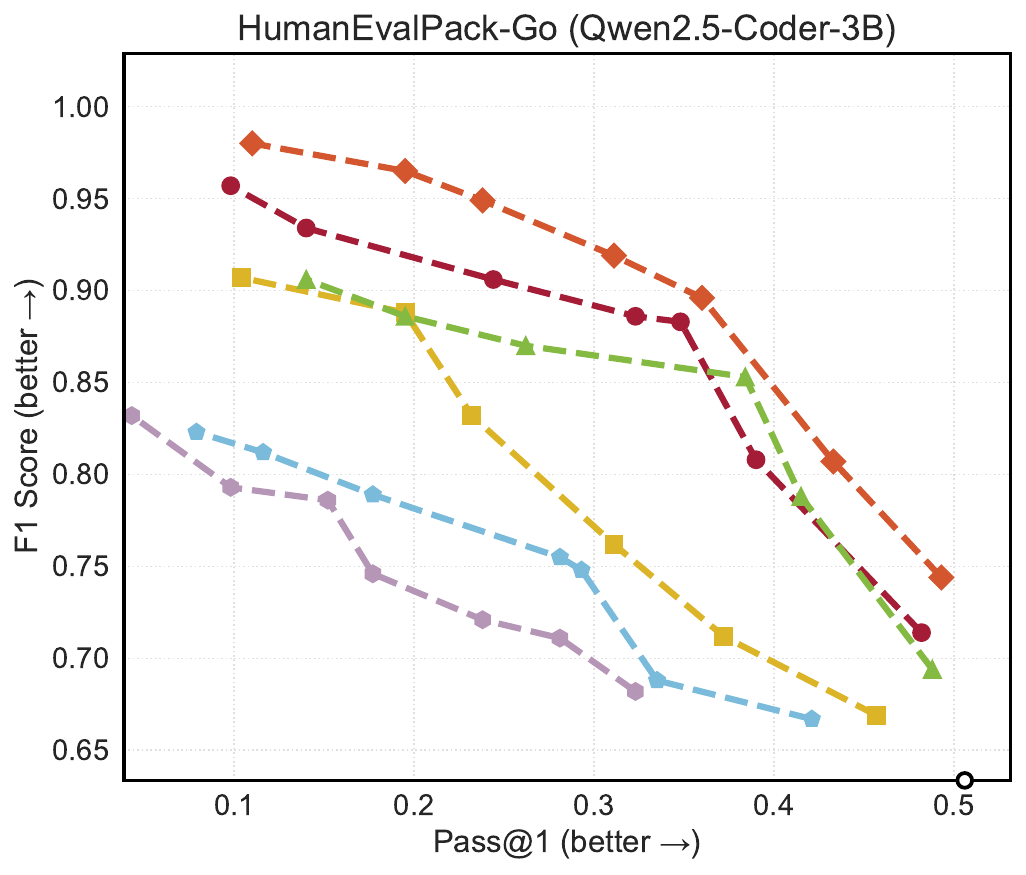}
\end{subfigure}
\hfill
\begin{subfigure}[t]{0.24\textwidth}
    \centering
    \includegraphics[width=\linewidth]{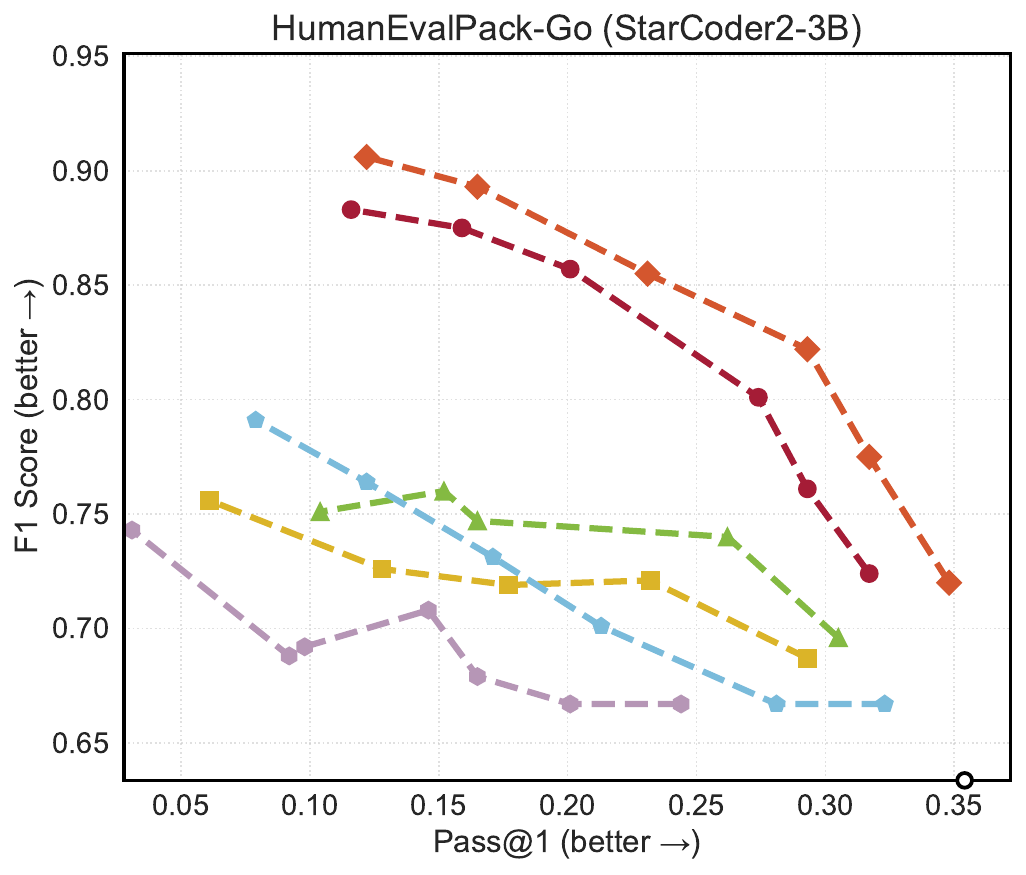}
\end{subfigure}

\begin{center}
\scriptsize
\setlength{\tabcolsep}{8pt}
\newcommand{\legendwithline}[3]{
\textcolor[HTML]{#1}{
\tikz[baseline=-0.6ex]{
\draw[line width=1pt] (0,0) -- (2.5em,0);
\node at (1.25em,0) {#2};
}}
\hspace{0.2em} #3
}
\begin{tabular}{ccccccc}
\legendwithline{A51C36}{$\bullet$}{KGW}
&
\legendwithline{DBB428}{$\blacksquare$}{SWEET}
&
\legendwithline{84BA42}{$\blacktriangle$}{EWD}
&
\legendwithline{6F6F6F}{$\bigstar$}{STONE}
&
\legendwithline{7ABBDB}{
\tikz\filldraw (90:0.7ex)
-- (18:0.7ex)
-- (-54:0.7ex)
-- (-126:0.7ex)
-- (162:0.7ex)
-- cycle;
}{CodeIP}
&
\legendwithline{B696B6}{
\tikz\filldraw
(0:0.7ex)
-- (60:0.7ex)
-- (120:0.7ex)
-- (180:0.7ex)
-- (240:0.7ex)
-- (300:0.7ex)
-- cycle;
}{SynthID-Text}
&
\legendwithline{D4562E}{$\blacklozenge$}{GDW (Ours)}
\end{tabular}
\end{center}

\caption{Pass@1-F1 quality-detectability trade-off curves on HumanEvalPack for Qwen2.5-Coder-3B and StarCoder2-3B under sampling (top row) and beam search (bottom row). The hollow marker represents the unwatermarked pass@1 performance as a quality reference. Curves closer to the upper-right region indicate a more favorable trade-off. }

\label{fig:HumanEvalPack_Trade-off}
\end{figure*}

We supplement the corresponding Pass@1-F1 quality-detectability trade-off curves on the HumanEvalPack benchmark, as shown in Figure~\ref{fig:HumanEvalPack_Trade-off}. Consistent with the observations in the main text, the curves of GDW are positioned above those of the baselines, indicating that GDW maintains favorable quality-detectability trade-off across programming languages Java and Go.

\section{Effect of Decoding Strategy: Beam Search vs. Sampling}

\begin{figure*}[t]
\centering

\begin{minipage}[t]{0.24\textwidth}
    \centering
    \includegraphics[width=\linewidth]
    {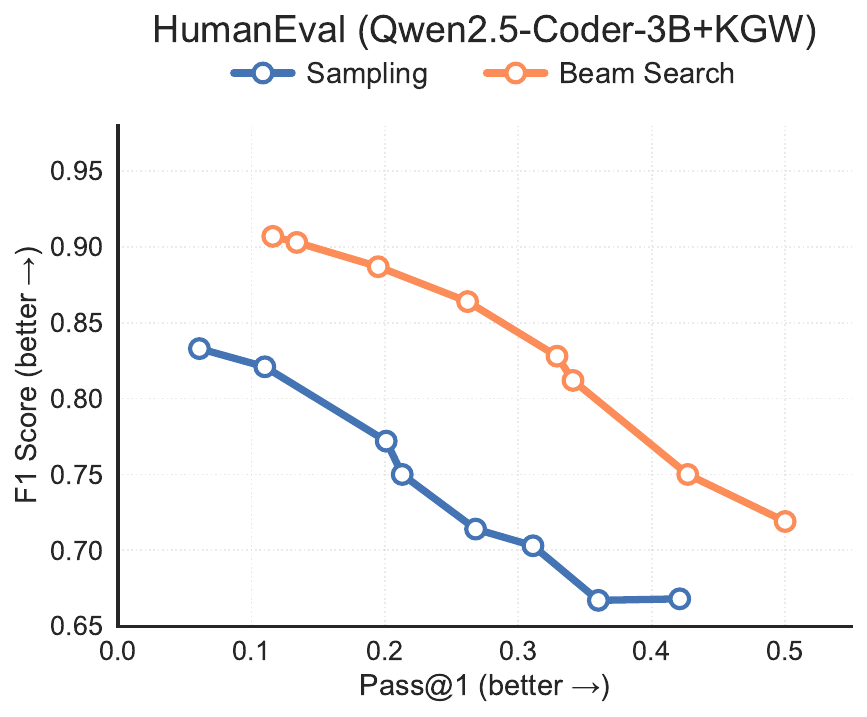}

    \includegraphics[width=\linewidth]
    {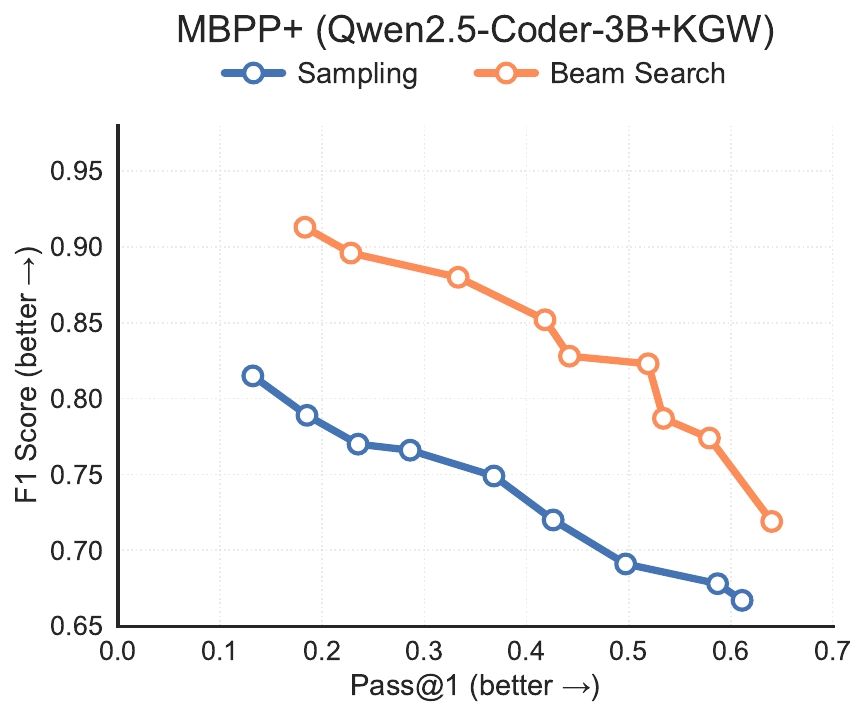}
\end{minipage}
\hfill
\begin{minipage}[t]{0.24\textwidth}
    \centering
    \includegraphics[width=\linewidth]
    {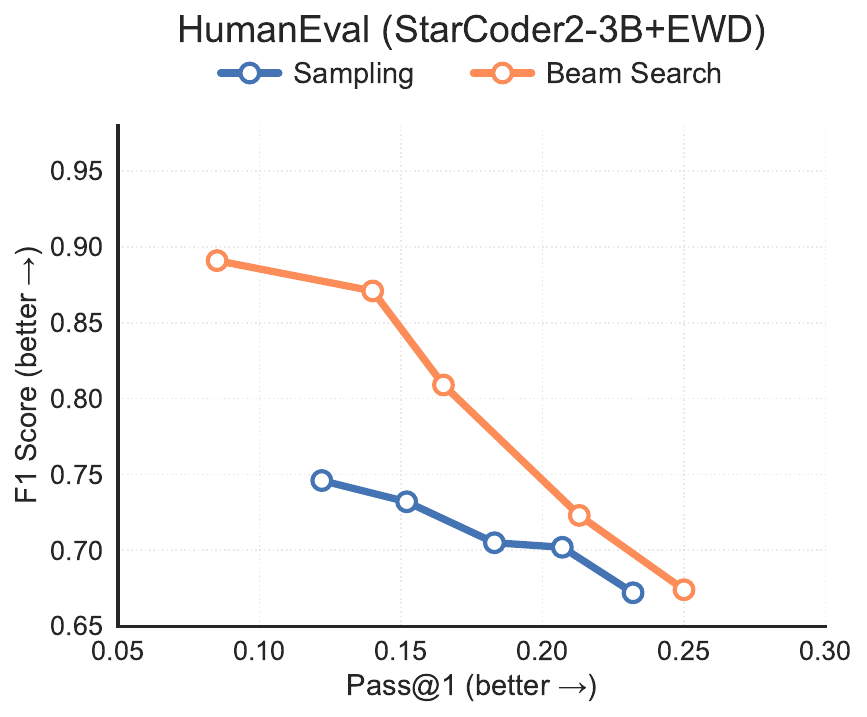}

    \includegraphics[width=\linewidth]
    {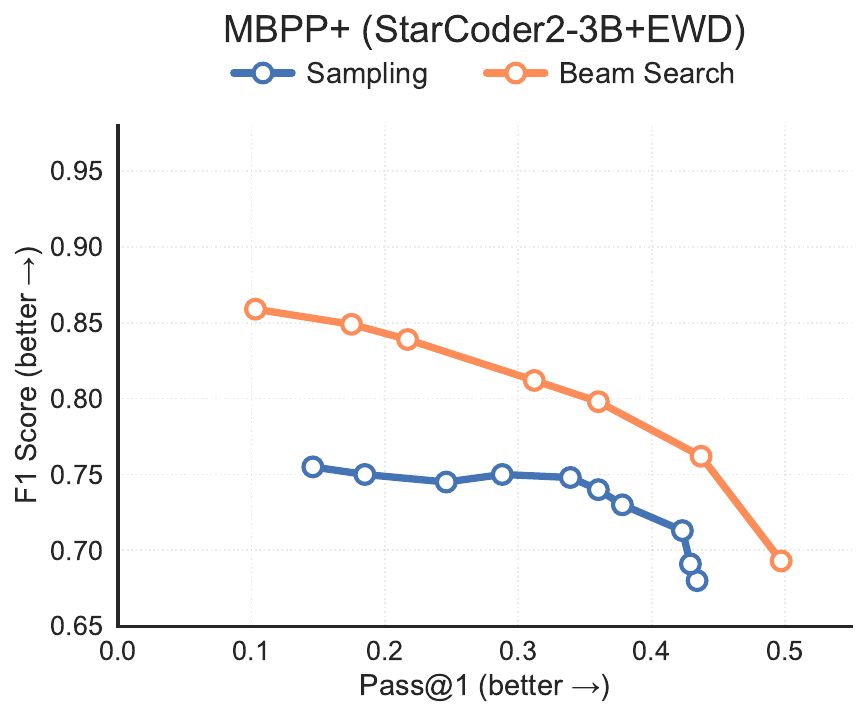}
\end{minipage}
\hfill
\begin{minipage}[t]{0.24\textwidth}
    \centering
    \includegraphics[width=\linewidth]
    {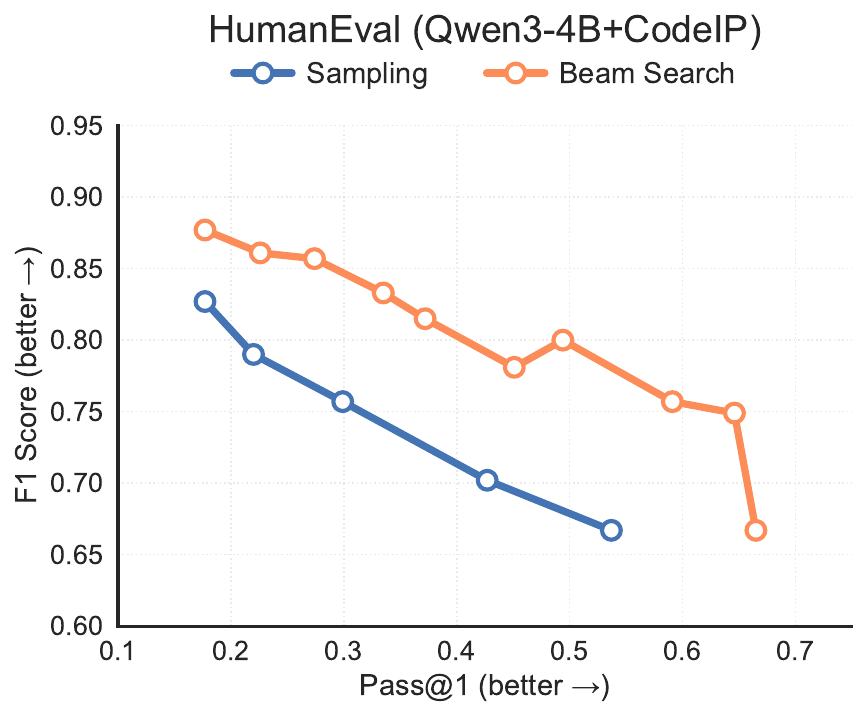}

    \includegraphics[width=\linewidth]
    {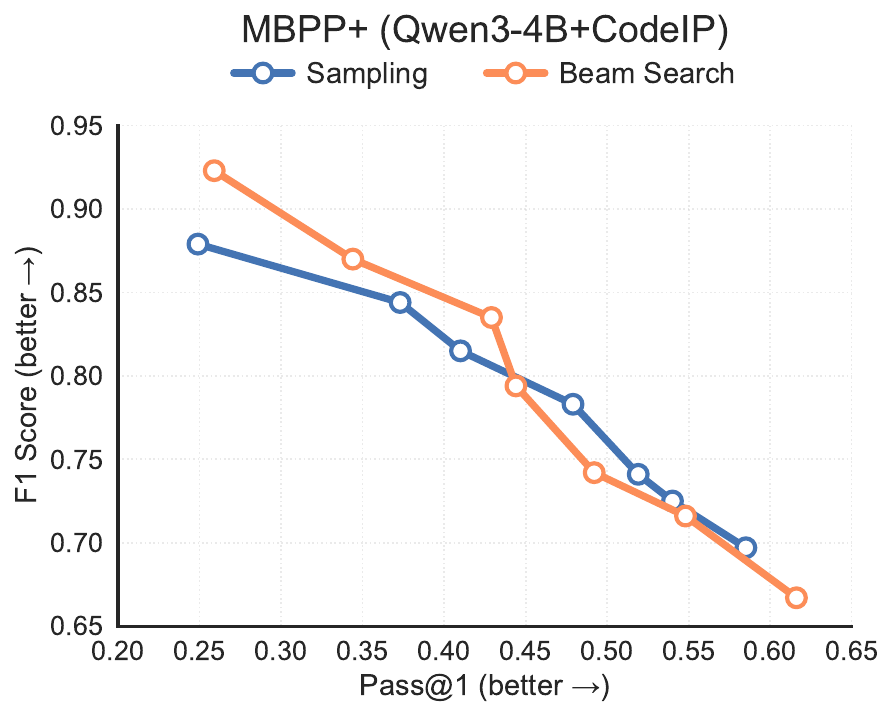}
\end{minipage}
\hfill
\begin{minipage}[t]{0.24\textwidth}
    \centering
    \includegraphics[width=\linewidth]
    {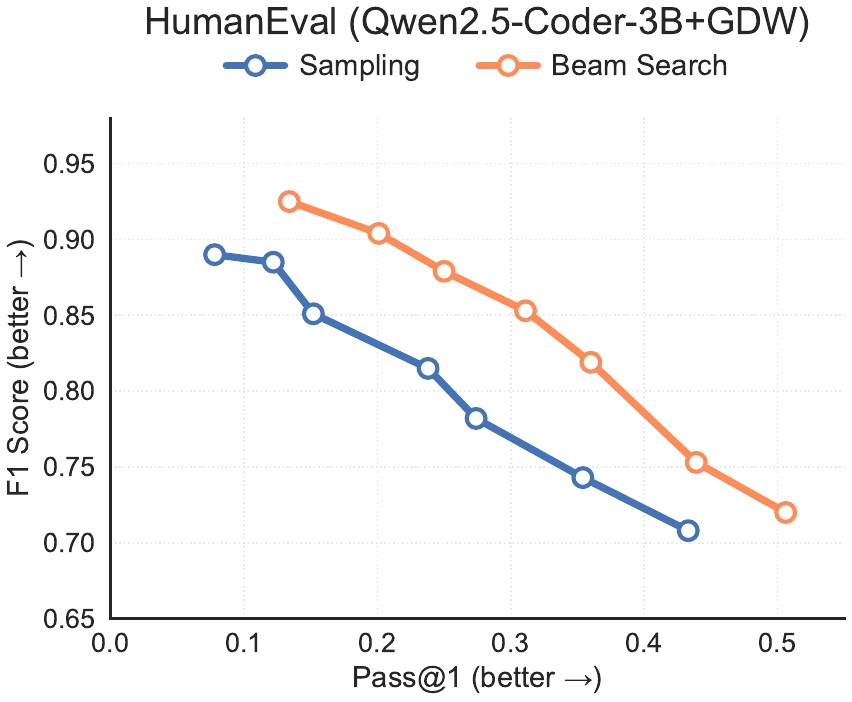}

    \includegraphics[width=\linewidth]
    {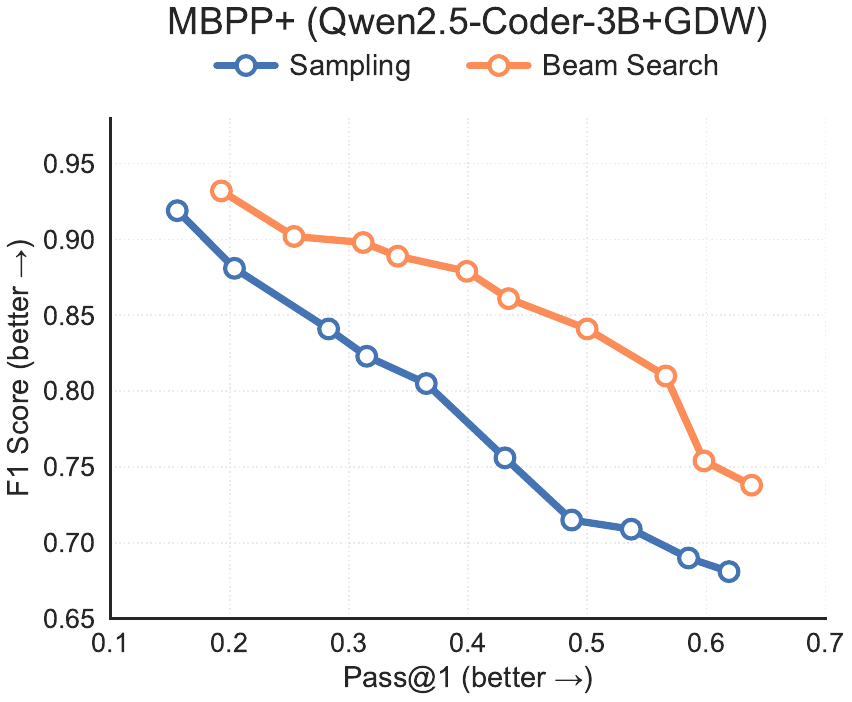}
\end{minipage}

\caption{Pass@1-F1 quality-detectability trade-off curves under sampling and beam search decoding. For conciseness, representative experimental settings are reported. The same trend, with beam search achieving a more favorable quality-detectability trade-off, is observed in the remaining settings.}
\label{fig:decoding_strategy_comp_all}
\end{figure*}

We compare nucleus sampling and beam search to examine how decoding affects watermark behavior. Figure~\ref{fig:decoding_strategy_comp_all} compares quality-detectability trade-off curves under beam search and nucleus sampling for eight representative experimental settings. In these settings, beam search generally shifts the curves toward the upper-right region, indicating a more favorable quality-detectability trade-off. We observe the same qualitative pattern in the remaining evaluated settings. One possible explanation is that code generation often exhibits concentrated probability distributions, where high-probability tokens are more likely to correspond to syntactically valid continuations, making deterministic decoding more likely to preserve high-probability generation paths and avoid quality degradation.

\end{document}